\documentclass[12pt,a4paper,div=11]{scrartcl}
\usepackage[T1]{fontenc}
\usepackage[utf8]{inputenc}
\usepackage[english]{babel}
\usepackage{amsmath,amsfonts,amssymb,amsthm}	% math environment
\usepackage{graphicx}
\usepackage{rotating}	% for sidewaystable
\usepackage{pdflscape}
\usepackage{subcaption}
\usepackage{multicol,multirow,booktabs,array,longtable}	% for making tables
\usepackage[flushleft]{threeparttable}
\usepackage{setspace}
\usepackage{xr}
\usepackage{url}
\usepackage{natbib}		% citation style
\usepackage{dcolumn}
\usepackage[hidelinks]{hyperref} % Clickable Links
\usepackage{color}
\usepackage{lmodern}
\setkomafont{caption}{\normalsize} % caption have normal size even in footnotesize longtable environment
\usepackage{ulem}
\usepackage{cancel}
\usepackage{soul} %Strike out paragraph
\usepackage{tabularx}

\newcommand{\sym}[1]{\rlap{#1}} % for stars in regression tables
\usepackage{siunitx}
\def\yyy{%
	\bgroup\uccode`\~\expandafter`\string-%
	\uppercase{\egroup\edef~{\noexpand\text{\llap{\textendash}\relax}}}%
	\mathcode\expandafter`\string-"8000 }
\def\xxxl#1{%
	\bgroup\uccode`\~\expandafter`\string#1%
	\uppercase{\egroup\edef~{\noexpand\text{\noexpand\llap{\string#1}}}}%
	\mathcode\expandafter`\string#1"8000 }
\def\xxxr#1{%
	\bgroup\uccode`\~\expandafter`\string#1%
	\uppercase{\egroup\edef~{\noexpand\text{\noexpand\rlap{\string#1}}}}%
	\mathcode\expandafter`\string#1"8000 }

\makeatletter
\edef\originalbmathcode{%
    \noexpand\mathchardef\noexpand\@tempa\the\mathcode`\(\relax}
\def\resetMathstrut@{%
  \setbox\z@\hbox{%
    \originalbmathcode
    \def\@tempb##1"##2##3{\the\textfont"##3\char"}%
    \expandafter\@tempb\meaning\@tempa \relax
  }%
  \ht\Mathstrutbox@\ht\z@ \dp\Mathstrutbox@\dp\z@
}
\makeatother

\usepackage{ulem}
\usepackage{xcolor}

\begin{document}
\title{Speed, Quality, and the Optimal Timing of Complex Decisions: \\ Field Evidence
\thanks{%The authors thank K.~Anders Ericsson, Armin Falk, Daniel Goller, Alex Krumer, Matthias Lang, David Smerdon, Andreas Steinmayr, and audiences at the CRC TRR 190 Meeting in Tutzing for useful comments.
Uwe Sunde gratefully acknowledges financial support of the Deutsche Forschungsgemeinschaft (CRC TRR 190 Rationality and Competition, project number 280092119). Anthony Strittmatter gratefully acknowledges financial support of the French National Research Agency (LabEx Ecodec/ANR-11-LABX-0047). \newline
Contact: Anthony Strittmatter anthony.strittmatter@ensae.fr, Uwe Sunde uwe.sunde@lmu.de, Dainis Zegners zegners@rsm.nl
}
}

\author{
\normalsize \textsc{Uwe Sunde} \\
%EndAName
{\small LMU Munich} \\
{\small CEPR, Ifo, IZA}
\and
\normalsize \textsc{Dainis Zegners} \\
%EndAName
{\small  Rotterdam}\\
{\small School of Management,}\\
{\small Erasmus University}
\and
\normalsize \textsc{Anthony Strittmatter} \\
%EndAName
{\small CREST-ENSAE, Paris}\\
{\small CESifo}
}

\date{\normalsize \today }
\maketitle

\begin{abstract}
\bigskip
This paper presents an empirical investigation of the relation between decision speed and decision quality for a real-world setting of cognitively-demanding decisions in which the timing of decisions is endogenous: professional chess. Move-by-move data provide exceptionally detailed and precise information about decision times and decision quality, based on a comparison of actual decisions to a computational benchmark of best moves constructed using the artificial intelligence of a chess engine. The results reveal that faster decisions are associated with better performance. The findings are consistent with the predictions of procedural decision models like drift-diffusion-models in which decision makers sequentially acquire information about decision alternatives with uncertain valuations.

%This paper presents novel evidence for the prevalence of deviations from rational behavior in human decisions -- and for the corresponding causes and consequences. Based on move-by-move data from chess tournaments, the identification strategy compares behavior of professional chess players to a rational behavioral benchmark that is constructed using modern chess engines. The evidence documents the existence of several distinct dimensions in which human players deviate from the rational benchmark and show that deviations do not necessarily lead to worse performance. Consistent with an influence of intuition and experience, faster decisions are associated with more frequent deviations and better performance.

\begin{description}
\item JEL-classification: D01, D9, C7, C8 \smallskip

\item Keywords: Response Times, Speed-Performance Profile, Drift-Diffusion Model, Uncertain Evaluations
\end{description}
\end{abstract}

\thispagestyle{empty}

\bigskip \thispagestyle{empty} \setcounter{page}{0} \normalem \newpage
\normalsize \renewcommand{\baselinestretch}{1.5}\normalsize

%\begin{quote}
%``%When you sit down to play a game you should think only about the position, but not about the opponent.
%Whether chess is regarded as a science, or an art, or a sport, all the same: psychology bears no relation to it and only stands in the way of real chess.'' \hspace{3.3cm} (Jose Raul Capablanca)
%\end{quote}

%\begin{quote}
%``I don't believe in psychology. I believe in good moves.'' \hspace{0.3cm} (Bobby Fischer)
%\end{quote}

%==============================================================================%
\section{Introduction}\label{sec: intro}
%==============================================================================%

Many real-world decisions in economics and management involve both substantial complexity and time pressure. The question of whether slow decisions or fast decisions are associated with higher quality remains an open issue, however. The relation between the speed and the quality of decisions in settings in which the timing of choices is endogenous
is unclear because different mechanisms work in opposite directions. Taking more time to reach a decision can help to make a more thoughtful choice, implying that slower decisions are of higher quality. Alternatively, slow decisions might entail lower quality if they reflect problems that decision makers perceive as more difficult, in terms of higher complexity or less clear information about the implications of different decision alternatives. Moreover, fast decisions can be of high quality when the decision maker has a strong intuition about the right choice or relatively precise information early on during the decision-making process so that the value of additional cogitation or information acquisition is relatively low. Existing evidence on the relation between the endogenous timing of decisions and their quality is scarce and largely limited to stylized experiments that often relate to fairly simple choices.

This paper presents an empirical investigation of the relation between decision speed and decision quality for cognitively-demanding real-world decisions in a competitive environment with high stakes: professional chess. Conceptually, chess is ideally suited to investigate this relation due to the extraordinarily rich and precise information about the context in which decisions are made. In particular, in this setting the complexity and difficulty of decisions exhibits considerable variation that can be measured with high accuracy, and the timing of decisions is endogenous. The decision making process in chess can be viewed as the solution to a complex problem in which the decision maker is uncertain about the quality of a particular move and, due to time restrictions, faces costs for gathering more information by cogitating over the various decision alternatives to determine the optimal move. The choice problem can thus be characterized as the determination of a move by an optimal stopping rule in a setting with uncertain priors about the quality of alternative moves. Conceptually, this closely resembles the decision problems often faced by managers in practice. Professional chess thus represents a unique source of high-resolution data from a non-experimental setting about behavior of decision makers facing high stakes, who are observed repeatedly while solving complex choice problems under time pressure, and with varying difficulty.

The empirical analysis is based on information for decision times and move quality for more than 80,000 moves from almost 1,600 games between professional chess players. For each move, the configuration of pieces on the board is analyzed with a state-of-the-art chess engine, which provides a precise and objective estimate of the best possible move for a player facing this configuration. For each move, the speed of the decision made by the player is available at high resolution. The use of artificial intelligence built into chess engines allows constructing clean and objective measures for the quality of a particular decision and for other relevant aspects such as the difficulty of a given decision problem. This makes it possible to measure decision time and decision quality with high accuracy. Moreover, the use of move-by-move data implies that the analysis is able to reveal general patterns of decision-making while absorbing systematic heterogeneity across individual decision makers.

The estimation results document two sets of findings. First, regarding the determinants of decision speed, the results show that decisions are faster when the remaining time budget of a player is more constrained, implying greater time pressure and hence a greater marginal cost associated with additional cogitation. Likewise, decisions are slower if the problem is objectively more complex or if the uncertainty about the best decision is greater. Second, faster decisions are robustly associated with higher decision quality. In addition, and conditional on decision times, lower decision quality is associated with higher time pressure, greater complexity and greater uncertainty.

\paragraph{Contribution to the Literature.}
%Different decision models have been developed to address this issue, but overall, the evidence for their respective predictions and the underlying mechanisms is scarce and largely limited to stylized experiments. For simple choices (like choosing between two different candy bars), fast decisions are often found to exhibit higher quality, as predicted by drift-diffusion models. For complex decisions (like choosing a strategy in a game), the prediction that slow decisions exhibit higher quality has found some empirical support.  %, which has been rationalized with two-system models of decision making.

The first set of findings corresponds to the intuitive predictions of models of decision making that
decision makers spend more time, which is a proxy for the effort devoted to cogitation, when the benefits of cogitation exceed
its costs \citep[see, e.g.,][]{gabaix2006,alaoui2016}. By considering field data with within-subject variation in decision times
conditional on fine-grained variation in time pressure, complexity, and uncertainty, our findings complement earlier lab findings of longer consideration
times when the gap in the subjective evaluation between two options is relatively small \citep{Chabris/etal:2009},
findings pointing at the sensitivity of reaction time with respect to strength of preferences
\citep[see, e.g.,][]{krajbich/etal:2015,konovalov2019}, recent work on deliberation times as correlate of cognitive effort influencing the depth and the intensity of reasoning \citep{alosferrer2021}, and work on adaptive decision making in strategic environments \citep{spiliopoulos2018b}.

Our second set of findings contributes to the small literature on the relation between endogenous decision times and decision performance, and on the underlying processes of reasoning in strategic decision contexts \citep[e.g.,][for a recent survey]{spiliopoulos2018a}. Response times have been studied based on notions of two-system models distinguishing between instinctive and cognitive (or contemplative) decision-making \citep{aradrubinstein2012,spiliopoulos2018b}, and for classifying individuals into instinctive and contemplative types \citep{rubinstein:2007,rubinstein:2016}. While this work typically focuses on observations of behavior of one individual in different games, our data comprise many observations for decisions of the same individual in the same decision context, but with varying complexity and time pressure. Only few studies have used repeated strategic interactions in the lab to measure the strategic complexity of a decision, with findings suggesting that decision times are responsive to strategic complexity and suggest that ``overthinking'' (i.e., thinking longer than usual) leads to lower decision quality when relating decision quality to decision times relative to decision times in similar decision settings  \citep[see, e.g.,][for a recent contribution using a $p$-beauty contest with feedback]{gill2019}. Our finding of a negative relation between decision time and decision quality is overall consistent with this, but our results are based on within-subject comparisons using variation in the complexity of decisions and in time pressure. By explicitly accounting for variation in difficulty and time pressure, our analysis addresses concerns about these as potential confounders in the respective literature \citep[see, e.g.,][for a discussion of this point and the need for incorporating difficulty]{krajbich/etal:2015,spiliopoulos2018a}.

Our findings are also informative about the underlying process of decision making. In particular, the negative relation between the time spent on the decision and the quality of the decision is inconsistent with simple two-system models of decision making that predict that faster decisions due to, e.g., greater time pressure and lower subjective importance, are less likely to be based on deep reasoning and thus exhibit lower quality in complex settings \citep[see, e.g.,][]{kahneman2003,Dane/Pratt:2007}. Instead, the negative correlation is consistent with two-system models in which decision speed and quality depend on the alignment of the solutions based on an intuitive system of decision-making (that might be based on experience accumulated in the past) and cogitation \citep{achtziger_alosferrer2016,Sahm/Weizsaecker:2016,alosferrer2018,spiliopoulos2018c}.
Earlier work has predicted that a greater time budget leads to better decisions by allowing for processing more decision alternatives \citep[see, e.g.,][]{busemeyer:1993}. Our findings are consistent with this prediction and confirm findings that additional time available for deliberation improves performance \citep[see, e.g.,][]{chabris2003,moxley2012}. However, our results also indicate the need to account for decision times when analyzing the role of time pressure on decision quality, since the positive relation between remaining time and decision quality only holds once conditioning on decision time. This is consistent with recent findings that the observed distribution of choices and their correlates critically depends on controlling for response time \citep{Webb:2019}. %Together with the negative relation between decision time and  decision quality, our findings suggest that the time budget serves as a proxy for additional behavioral mechanisms, such as stress.
Our results allow disentangling the roles of available time budget (which reflects the implicit costs for later decisions), and the complexity and uncertainty associated with the decision, for the speed and quality of the decision.

Our empirical results are also consistent with the predictions of recent work on drift-diffusion models that has considered settings in which the relative evaluations of decision alternatives are uncertain ex-ante and that has shown that the positive association between decision speed and decision quality dominates in such settings \citep[see][for details]{Fudenberg/etal:2018,Fudenberg/etal:2020}. So far, drift-diffusion models have been mainly used to model decisions in simple decision problems and existing evidence is restricted to lab experiments with non-complex decisions \citep[see, e.g.,][]{Ratcliff/Rouder:1998,krajbich2011,Ratcliff/etal:2015,Bhui:2019a,Schotter/Trevino:2021,Huseynov/Palma:2021}. Our results document that the relevance of drift-diffusion models with uncertain evaluations of decision alternatives extends to complex decision contexts. Moreover, our findings provide evidence consistent with the underlying mechanisms. To our knowledge, the present study is the first to provide evidence from a realistic field setting with experienced decision makers and an objective measure of decision quality.
On a more applied side, our results inform the literature on intuitive decision making by experts \citep[see, e.g.,][for a review]{Salas/etal:2010}. In this context, our results complement empirical work on bounded rationality \citep{Zegners/Sunde/Strittmatter:2020b}, suggesting that human intuition and experience are important factors in determining optimal decisions.

Our analysis continues with some conceptual considerations and a description of the data and the empirical measures in Section \ref{sec: implementation}. Section \ref{sec: results} contains the results. In Section \ref{sec: conclusion}, we provide a discussion of the results and show that the negative relation between time spent on a decision and the quality of the decision is also consistent with the predictions of drift-diffusion-models.

%==============================================================================%
\section{Empirical Implementation \label{sec: implementation}}
%==============================================================================%

%==============================================================================%
\subsection{Conceptual Considerations  \label{sec: hypotheses}}
%==============================================================================%

The analysis of the relationship between decision speed and decision quality is rooted in a conceptual setting in which a decision maker
is confronted with several decision alternatives and faces time pressure. Concretely, the decisions in the empirical setting of chess involve making a decision about the next move in a particular configuration of pieces on the board under a limited time budget. Decisions involve the choice of a particular move out of a set of alternative moves, which corresponds to a discrete choice problem. The underlying neural dynamics of decision making imply that decision speed and decision quality are closely related. An approach to model these decision dynamics used in bounded accumulation models from psychology is to assume that the cogitation process corresponds to the sequential acquisition and processing of information about the consequences of different move alternatives, and an optimal stopping rule. This implies that the observed distribution of decisions and their respective quality crucially depends on decision times.

Empirical assessments of the relation between decision times and decision quality typically focus on particular regularities such as individual differences in response time and accuracy, often using between-subject designs or within-subject designs with only few choices. This raises identification problems related to unobserved heterogeneity across decision makers and decision problems. The ideal setting to investigate the relationship between decision quality and decision time involves a within-subject design with many observations of similar, but not identical, decisions. The chess data used in the empirical analysis below provide such a setting of within-subject variation with observations of a large number of decisions and the associated decision speed and quality for the same decision maker. By their nature, moves in chess constitute full information rational choice problems under a time constraint and with uncertainty about the best possible solution. Moreover, in chess, decision makers only rarely face the exact same choice problem, especially when abstracting from the opening phase of the game, due to the abundance of different configurations of pieces on the board, which creates considerable variation in time pressure and in the difficulty of the decision. Hence, the empirical implementation is based on information about the decisions of the same individual across choice problems that differ due to the difference in the configuration on the board.

Our analysis proceeds in two steps. The first step refers to the factors that influence decision times, and the second step considers the relation between the quality of decisions and decision times as well as the other decision-relevant factors. In the first step, we consider time pressure and the difficulty of the decision as main determinants of decision times. The testable hypotheses are evaluated using proxies for these determinants. In particular, as a proxy for time pressure, we use the remaining time budget for a given player in a given configuration of the game. A smaller time budget implies a larger opportunity cost since spending an additional unit of time on cogitating over a move implies less time available for future moves. This constraint becomes increasingly binding as the remaining time budget converges to zero. Hence, greater time pressure (in terms of a smaller remaining time budget) implies higher costs for later decisions in the game and should therefore lead to faster decisions.  Coming up with proxies for the difficulty of the decision problem is more complicated since difficulty can arise for distinct reasons. In particular, greater difficulty can arise from greater positional complexity, where more complex problems imply a slower process until a decision can be reached with sufficient confidence. This can be either due to a slower process of cogitation and information acquisition, or due to a more noisy and unpredictable cogitation process. As a proxy for this feature, we use a measure of complexity of a given configuration that is based on the number of alternative moves that need to be considered in a given positional setting. Alternatively, difficulty can arise from greater uncertainty related to the evaluation of alternatives. This affects the possibility to discriminate between alternatives and hence the uncertainty associated with a decision. To capture this dimension, we use information about the evaluation gap between the best conceivable move for a given configuration, and the second-best move. Greater difficulty, either in terms of greater complexity of the decision or in terms of a more diffuse prior about the optimal decision, is expected to lead to longer decision times until a choice is made.

In the second step, we investigate the relation between decision quality and decision times, as well as the role of the factors that influence decision times. The relation between decision quality and decision times in complex, cognitively-demanding tasks is an empirical question and depending on the particular model, different predictions are possible. A similar statement applies to the influence of time pressure and the difficulty of the decision. It is likely, however, that the influence of these factors on decision quality crucially depends on the timing of decisions.

%In sum, the hypotheses that follow from these considerations refer to decision time and decision quality. Everything else equal, decision makers are expected to spend more time on a decision if the remaining time budget is larger; if the problem is more difficult in terms of complexity of a given configuration; and if the uncertainty about the evaluation of decision alternatives is larger, i.e., the smaller the gap in the evaluation between the best move and the second-best move. In terms of decision quality, the predictions are less clear and the relation between decision-relevant factors and decision quality is likely to depend on the speed of decisions.

%==============================================================================%
\subsection{Data and Measurement}\label{sec: data}
%==============================================================================%
\paragraph{Data from Professional Chess Players.}
%In the terminology of game theory, chess is a two-person, sequential, zero-sum game with perfect information and alternating moves, for which the optimal strategy is strictly determined.\footnote{See \citet{Schwalbe/Walker:2001} for details and a discussion of the historical background.}
The data used in the empirical analysis have been collected from an internet platform that broadcasts all professional over-the-board chess tournaments (\url{www.chess24.com}) and contains detailed information for more than 80,000 moves from around 1,600 games that were played in 54 single round-robin tournaments during the years 2014-2017. These include events such as national championships, qualifications tournaments for the world championship and the most prestigious invitational tournaments in the chess calendar. Table \ref{tab: tournaments} in the Appendix provides a list of tournaments included in the data set. All games were played at regular time controls that allocate a time budget of a minimum of 2 hours thinking time to each player to conclude the game.\footnote{According to the regulations by the International Chess Federation FIDE, for a game to be rated for determining the world rankings each player must have a minimum of 120 minutes, assuming the game lasts 60 moves per player
%. The standard time control regime suggested by the International Chess Federation FIDE is 90 minutes per player per game plus 30 seconds added to each player's time budget for each move played; additional 30 minutes are added to each player's time budget after each player has played 40 moves
(see \url{https://handbook.fide.com}, last accessed January 3, 2022). %\url{https://handbook.fide.com/chapter/B022017}
Tournaments that are not officially organized by FIDE use slight variations of the official FIDE time control regime.
%Appendix Table \ref{tab:tournaments} provides an overview of the tournaments contained in the data set.
}
%On average, the data contain 10 players and 44 games per tournament.
The data contain detailed information about the players, including their performance statistics in terms of their ELO number.\footnote{The ELO number constitutes a method for calculating the relative playing strength of players (invented by the Hungarian mathematician Arpad Elo).
%The ELO number increases or decreases depending on the outcome of games between rated players. After every game, the winning player takes points from the losing player, while the total number of points remains fixed.
According to international conventions, an ELO number of at least 2,500 is a requirement for being awarded the title of an international grandmaster (this requirement has to be fulfilled once during the career, but does not have to be maintained to keep the title, see \texttt{https://handbook.fide.com/chapter/B01Regulations2017}, last accessed January 3, 2022).}
%2017 eight players in the world had ratings with ELO$>$2800 (Carlsen, Anand, Topalov, Nakamura, Kasparov, Caruana, So, Kramnik).}
%The difference in the ELO-ratings between two players can be seen as an (ex-ante) predictor of the likely outcome of a game.
We restrict our analyses to games between professional players with an ELO number of at least 2,500 at the time of the game.\footnote{We also excluded games with decision times that are likely to be mismeasured (such as negative decision times or times of several hours devoted to a single move) and all games from tournaments that included multiple games exhibiting problematic decision time data.}

%Appendix Table \ref{tab: descriptive statistics} shows summary statistics on the move level. (Dainis: Auf die Summary stats wird später nocheinmal verwiesen darum nehm ich das hier raus)
%\footnote{Details about player strength and nationality are contained in Tables \ref{tab: player overview} and \ref{tab: player nationality overview} in the Appendix.}

The data include precise information about the time consumed for each move and about the remaining time budget. In addition, the move-by-move data comprise information about the exact configuration of pieces on the board. We use this information to compute an evaluation of this configuration in terms of the relative standing of each player, an evaluation of the complexity of the configuration, and an evaluation of move quality, using a chess engine as explained in more detail below. We exclude the first fifteen moves of each player in a game from all our analyses. These moves are typically the result of routine openings, which are studied intensively and memorized by players in the preparation for a game.

\paragraph{Measuring Decision Quality.} To construct a measure of decision quality, we make use of a chess engine that computes the best possible move for a given configuration of pieces on the chessboard. In particular, we use the open-source chess engine \textsc{Stockfish}, which is considered to be one of the best available programs and has an estimated ELO rating of approximately 3544 points.\footnote{We use the version \textsc{Stockfish 14} with a strength of play that is equivalent to ELO of 3544 points according to \url{https://ccrl.chessdom.com/ccrl/4040/} (accessed on December 20, 2021).
%\textcolor{blue}{\sout{According to} \cite{ferreira2013}\sout{, reducing the search depth by one move reduces an engine's playing strength by about 66 ELO points. Ferreira estimates this relationship using the chess engine \textsc{Houdini} which he estimates to have a chess strength of ELO 2894 at a search depth of 20 moves. As \textsc{Stockfish 8} in its unconstrained version is estimated according to CCRL to have an ELO rating 191 points higher than \textsc{Houdini}, we estimate the ELO strength of \textsc{Stockfish 8} at a search depth of 21 as 2894 + 191 + 66 = 3151 (see \href{https://web.archive.org/web/20190910145801/http://ccrl.chessdom.com/ccrl/4040/cgi/compare_engines.cgi?family=Houdini&print=Rating+list&print=Results+table&print=LOS+table&print=Ponder+hit+table&print=Eval+difference+table&print=Comopp+gamenum+table&print=Overlap+table&print=Score+with+common+opponents}{http://ccrl.chessdom.com}, archived on September 10, 2019). In comparison, Magnus Carlsen's ELO number in January 2020 was 2872. Source: \url{https://ratings.fide.com/toparc.phtml?cod=577}, last visited March 10, 2020.}}
} In comparison, the incumbent World Champion Magnus Carlsen had an ELO rating of 2856 points in December 2021 according to the official rating list by the International Chess Federation FIDE.
%This engine can be restricted, such that the strength of play of the engine corresponds closer to the strength of play of human players. To find an objective benchmark for human players, we use engines with different strengths of play in the analysis as described in detail below.

%An engine behaves exactly like stipulated by standard game-theoretic considerations.
%Figure \ref{fig: game tree} illustrates the decision algorithm solved by the engine.
For each configuration, the chess engine creates a game-tree for all possible moves by white and black for a pre-specified depth of $n$ moves ahead, the so-called search depth. We restrict the engine to a search depth of 24 moves to economize on computing costs. Then, the configurations at the respective end-nodes are evaluated in terms of pieces left on the board, safety of the king, mobility of pieces, pawn-structure, etc. Based on this evaluation, the engine then determines the best move using an algorithm conceptually similar to backward induction under the assumption of mutually best responses.\footnote{Most modern chess engines are based on domain-specific algorithmic heuristics that were developed specifically to search the sequential game-tree arising from a given configuration. In particular, these chess engines use an enhanced version of the min-max algorithm that disregards branches of the search tree that have already been found to be dominated. This reduces the search-space without impacting the final choice of the best move by the engine (\url{https://www.chessprogramming.org/Alpha-Beta}, last visited April 18, 2021). Only recently, more general machine learning techniques in the form of neural networks have been embodied in chess engines such as Google's non-public engine \emph{AlphaZero} \citep{silver2018} and the open-source engine \emph{LeelaChessZero}. We use a version of \emph{Stockfish 14} that searches the game tree using the standard approach of Alpha-Beta search and evaluates the end positions at the end nodes using a neural network trained on a large database of pre-evaluated chess positions  (\url{https://stockfishchess.org/blog/2020/introducing-nnue-evaluation}, last visited January 5, 2022). It is very likely that \emph{Stockfish 14} is considerably stronger than \emph{AlphaZero} based on the fact that Stockfish's current ELO is estimated to be by about 200 points higher than the version of Stockfish that was defeated by \emph{AlphaZero} in 2018 (see \url{https://chess.stackexchange.com/questions/29791/which-is-better-stockfish-10-or-alphazero} (accessed January 17, 2022)). Since Google's  \emph{AlphaZero} is non-public and has not played any publicly released games since 2018, this assessment cannot be validated, however.}

%\begin{figure}[]
%\begin{center}
%\includegraphics[width=0.7\textwidth]{Example_How_Engine_Thinks.pdf} \\
%  \caption{Backward Induction by Chess Engines}\label{fig: game tree}
%\vspace{0.2cm}
%\parbox{15cm}{%Panel (a):
% \footnotesize \emph{Note: } Illustration of the decision algorithm built into a chess engine. For a given search depth (number of moves until the end node is reached), the engine calculates evaluations of different alternative moves under the assumption of mutually best response and determines the move that delivers the highest evaluation on the end node.}
%\end{center}
%\end{figure}

Concretely, the chess engine delivers a measure of the relative standing for a given configuration of pieces on the board. This measure reflects an evaluation of the current position of a player and represents a proxy of the winning odds. The evaluation of the current position is the result of the engine computing, for each configuration observed in the data set, the best continuation, and the relative standing is measured in terms of so-called pawn units, where one unit approximates a continuation that is associated with an advantage of possessing one more pawn.\footnote{This measure is relative. Positive numbers indicate an advantage for the player with white pieces, negative numbers indicate an advantage for the player with black pieces. For example, if the evaluation is -1.00 pawn units, black is better ``as if one pawn up.''}
Based on this information, we compute a measure of performance, in terms of the quality of play of a given player in a given positional setting. This measure is based on the comparison of the actual move made by the respective player to the best conceivable move as suggested by the chess engine. This move is not necessarily the absolut best move, but on average the move suggested by the engine is better than conceivable by any human player. Decision quality is measured by a binary indicator of whether a player makes the best possible move as suggested by the chess engine in a given configuration.
%\footnote{Concretely, we configure the engine to compute the evaluations for the six moves that it evaluates as best in a given configuration. If the actual move played is one of these six moves, the performance is calculated as the difference in evaluation between the best and the actual move played. If the move played is not among the six best moves, we compute performance as the difference in the evaluation right before and right after the respective move of the player. Further increasing the number of moves that are evaluated comes at a prohibitively large computational cost.}
In robustness checks, we also computed the deviation of a player's move (in terms of pawn units) from the best move identified by the chess engine.

%\begin{figure}[]
%\begin{center}
%\includegraphics[width=0.9\textwidth]{Example_Output_Chess_Engine_Final}
%\caption{Computation of Performance Measures: An Example}\label{fig: evaluation example}
%\end{center}
%\footnotesize{\emph{Note: } The engine evaluates the configuration shown on the board as +0.95 (i.e., an advantage for white of almost one pawn unit) if Black plays Knight to b4 as the next move. Instead, Black played Queen takes b2, which the engine judges as a slight mistake, with the consequence of an evaluation of +1.14 for White after this move. Hence, the quality of Black's move is computed as -0.19, i.e., Black played a move that resulted in the loss of 0.19 pawn units compared to the evaluation resulting after the move suggested by the engine. In this example, the engine needed 3.87 seconds to reach a search-depth of 21 moves, which corresponds to the measure of complexity of the configuration.}
%\end{figure}

%Summary statistics on the level of individual moves are contained in Table \ref{tab: descriptive statistics} in the Appendix.

\paragraph{Proxy Variables of Decision-relevant Factors.} We use the available information to construct three proxy measures for factors that are central to the empirical analysis of the testable hypotheses.
%This gives a measure of the quality of a move by a player relative to the best conceivable move, measured in terms of pawn points.
First, we use the time budget in terms of remaining time as a proxy of the (inverse) time cost for cogitation. Second, we use a chess engine to compute, for each observed configuration, a measure of complexity of the configuration. This measure corresponds to a proxy for the difficulty of the decision. The more complex the configuration, the more nodes the chess engine needs to search in the game-tree arising from a given position. The number of nodes needed for the engine to compute the best strategy for the next $n$ moves ahead therefore serves as a measure of the complexity of a given configuration.\footnote{As a baseline measure of complexity, we use the number of nodes computed by the chess engine to reach a search-depth of 24 moves.} In particular, we use the number of \emph{mega-nodes} (number of \emph{nodes} dived by $1,000,000$) as a measure of complexity.\footnote{Modern engines like \textsc{Stockfish} calculate approximately 10-100 million nodes per second on standard personal computing hardware.}
%Figure \ref{fig: evaluation example} contains a concrete illustration of how these measures are computed.
Third, we compute a proxy of the (inverse of) uncertainty of the prior, in terms of the difference between the best continuation as stipulated by the chess engine and the second-best continuation. The greater this difference, the smaller the uncertainty about the relative evaluation of the best move and any other (i.e., the second-best) alternative; the smaller this difference, the more diffuse the prior (and the greater the uncertainty) about the best continuation move.
%\footnote{In particular, we restrict Stockfish 8 to a search depth of 12 moves, which corresponds to a play strength equivalent to an ELO of around 2700 when comparing performance differences between human players and the restricted engine.
%In a companion paper \citep{Strittmatter/Sunde/Zegners:2018b} we use the restriced engine as a benchmark for bounded rationality.}

%Appendix Figure \ref{fig: player strength and quality} shows that stronger players (better ranked in terms of their ELO number) are on average more likely to play the optimal move and deviate less in terms of pawn units from the optimal move suggested by the super chess engine, which validates the measure of quality of chess play.
Appendix Table \ref{tab: descriptive statistics} documents the descriptive statistics of the move-by-move data used in the analysis.

%==============================================================================%
\section{Empirical Results \label{sec: results}}
%==============================================================================%

\paragraph{Decision Speed.} The first set of results relates to the determinants of decision time. Table \ref{tab:time_spent} presents the results of multivariate regressions that include controls for (interacted) player-game fixed effects, the evaluation of the current position from the perspective of the decision maker, and for previously played moves (fixed effects for the number of moves played) to capture systematic heterogeneity that is not related to the research question.

\begin{table}[!t]
	\begin{center}
		\caption{Determinants of Decision Time \label{tab:time_spent}}
		\footnotesize
		{
\def\sym#1{\ifmmode^{#1}\else\(^{#1}\)\fi}
\begin{tabular*}{1\hsize}{@{\hskip\tabcolsep\extracolsep{-10pt}}l*{4}{D{.}{.}{-1}}}
\toprule
 &\multicolumn{4}{c}{Dependent Variable:}\\
                    &\multicolumn{4}{c}{Time Spent on Move (minutes)}                                       \\\cmidrule(lr){2-5}
                    &\multicolumn{1}{c}{(1)}         &\multicolumn{1}{c}{(2)}         &\multicolumn{1}{c}{(3)}         &\multicolumn{1}{c}{(4)}         \\
\midrule
\emph{Time Budget}  &                     &                     &                     &                     \\
Remaining time (min.)&      0.0985\sym{***}&                     &                     &      0.0947\sym{***}\\
                    &    (0.0032)         &                     &                     &    (0.0032)         \\
\emph{Complexity}   &                     &                     &                     &                     \\
N Mega-nodes computed&                     &      0.0336\sym{***}&                     &      0.0193\sym{***}\\
                    &                     &    (0.0031)         &                     &    (0.0028)         \\
\emph{Evaluation Gap}&                     &                     &                     &                     \\
Distance second best move (log)&                     &                     &     -0.9195\sym{***}&     -0.8294\sym{***}\\
                    &                     &                     &    (0.0241)         &    (0.0233)         \\
\midrule
Player-Game Fixed Effects&         Yes         &         Yes         &         Yes         &         Yes         \\
Number Move Fixed Effects&         Yes         &         Yes         &         Yes         &         Yes         \\
Control Evaluation Position&         Yes         &         Yes         &         Yes         &         Yes         \\
Move Observations   &       80359         &       80359         &       80359         &       80359         \\
Game Observations   &        1592         &        1592         &        1592         &        1592         \\
\bottomrule
\end{tabular*}
}

		\parbox{\textwidth}{\footnotesize \emph{Note: }{OLS estimates. The variable \emph{Distance second best move (log)} is computed as $ln(d+1)$ where $d$ is the absolute difference between the evaluation of the best and the second best move in terms of pawn units as given by the chess engine. The evaluation of the current position is controlled for using two dummy variables indicating whether the current position is evaluated as better ($>0.5$ pawn units) or worse ($<-0.5$ pawn units) for the player to move. Standard errors are clustered on the game level. $^{*}$: $p<0.1$, $^{**}$: $p<0.05$, $^{***}$: $p<0.01$. }}
	\end{center}
\end{table}

\begin{figure}[h!]
\begin{center}
\begin{subfigure}{0.45\textwidth}
\includegraphics[width=1\textwidth]{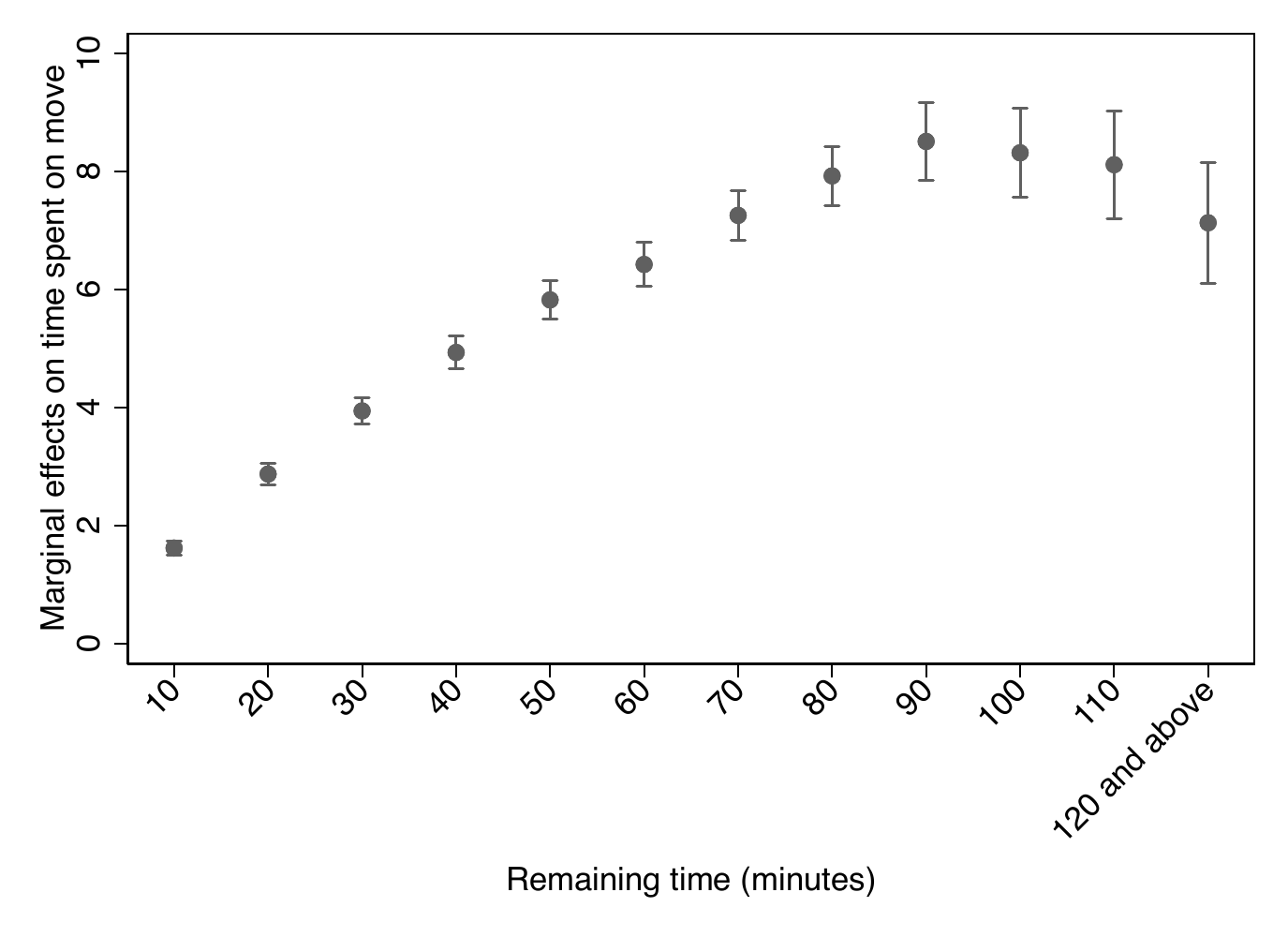}
\caption{Time budget}
\end{subfigure} \vspace{1cm}
\begin{subfigure}{0.45\textwidth}
\includegraphics[width=1\textwidth]{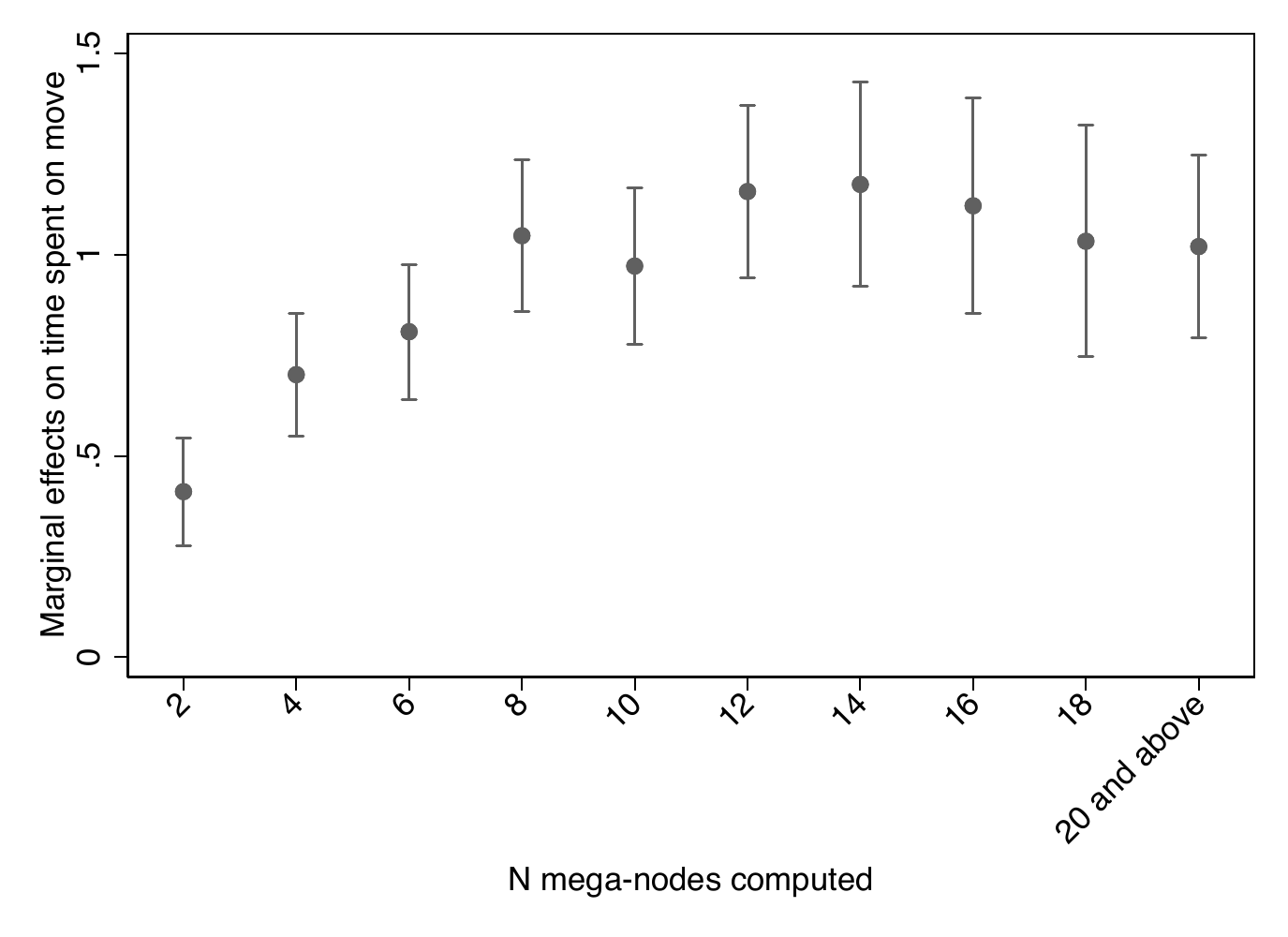}
\caption{Complexity}
\end{subfigure}
\begin{subfigure}{0.45\textwidth}
\includegraphics[width=1\textwidth]{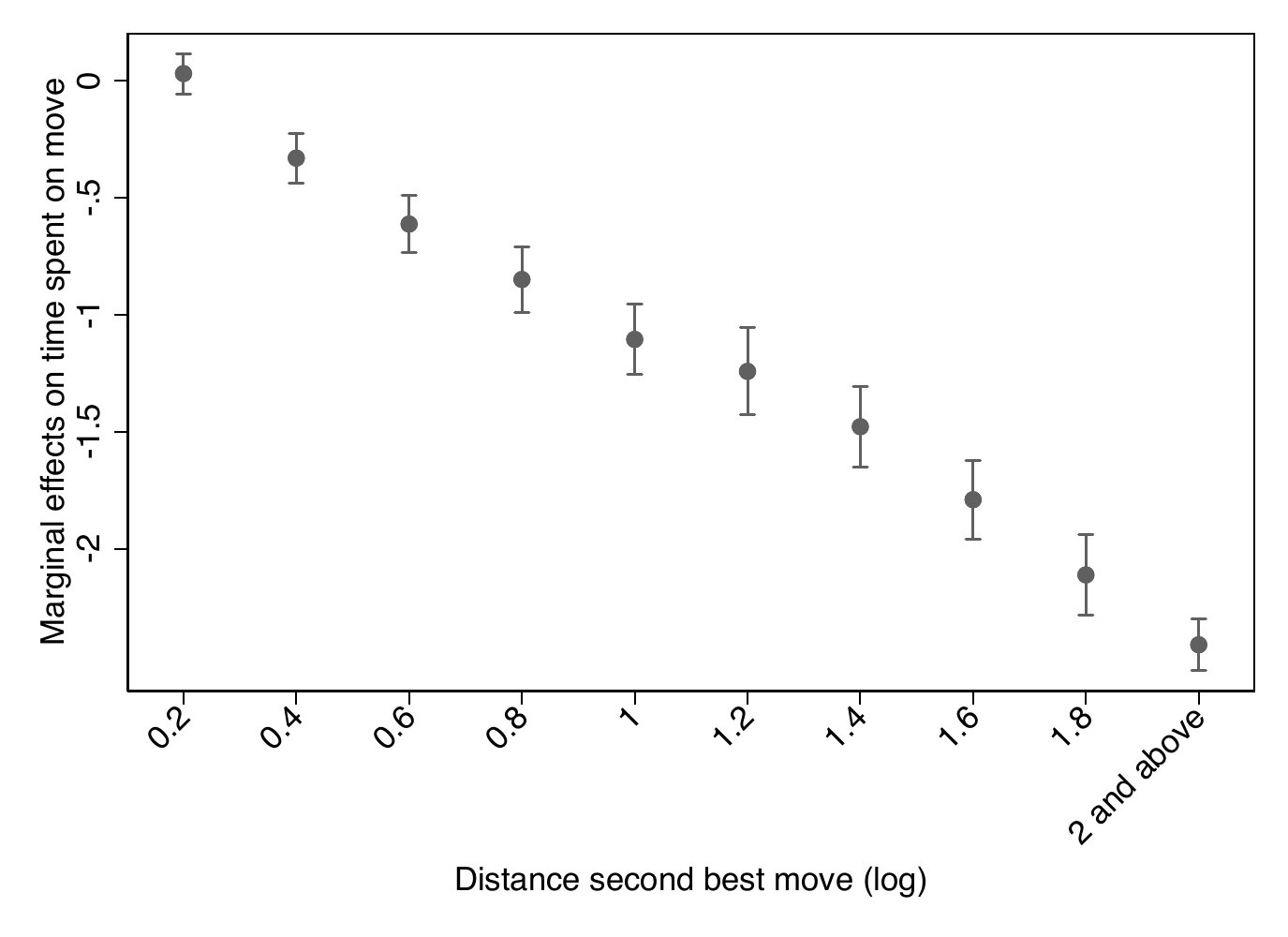}
\caption{Evaluation gap}
\end{subfigure} \\ \vspace{1cm}

\caption{Determinants of Decision Time - Flexible Specifications \label{fig:speed conditional}}
\vspace{0.2cm}
\parbox{15cm}{
	\footnotesize \emph{Note:}{ OLS results of flexible specifications. Dependent variable is Time Spent on Move (minutes). All three graphs are based on estimates of flexible dummy specifications of variables from the same regression model that includes the same controls and fixed effects as the specifications in Table \ref{tab:time_spent}.
Variables depicted on the horizontal axis are split into equally spaced intervals. Dots report point estimates and whiskers report 95\% confidence	intervals.
Panel (a): Remaining Time measured in minutes; omitted reference category is Remaining Time $<5$ minutes.
Panel (b): Complexity of chess position as measured by number of mega-nodes ($1$ mega-node equals $1,000,000$ nodes) computed by chess engine to reach search depth of 24 moves ahead; omitted reference category:  $<1$ mega-nodes computed.
Panel (c): Distance second best move is computed as $ln(d+1)$ where $d$ is the absolute difference between the
evaluation of the best and the second best move in terms of pawn units as given by the chess engine;
omitted reference category is distance second best move $<0.1$.
}
}
\end{center}
\end{figure}

The first finding, presented in Column (1), shows that decision makers spend more time on a move when the remaining time budget is larger. The results in Column (2) show that a greater complexity of a given configuration, as proxied by the number of mega-nodes computed by the chess engine to determine the best move for a given search depth, is associated with longer decision times. The results in Column (3) document that decision times are longer if the evaluation gap (i.e., the difference in the evaluation) of the best and of the second-best move is smaller. A smaller evaluation gap serves as proxy about greater uncertainty regarding the quality of alternative moves. These results also emerge when estimating a specification that includes all covariates, as shown in Column (4). The corresponding results for non-linear relationships between decision time and the respective measures are shown in Figure \ref{fig:speed conditional}. Considering more flexible specifications confirms the main findings: Greater time pressure,less complexity and less uncertainty all imply faster decisions. These findings could be due to higher costs for cogitating, or due a faster cogitation process that induce faster decisions, respectively.

These results are robust to using alternative measures for the proxy variables for time pressure, complexity and the uncertainty regarding the best move (see Table \ref{tab:time_spent_robust} in the Appendix).\footnote{In particular, as alternative proxy for time pressure we use an indicator for immediate time pressure characterizing moves before the time control at 40 moves. As an alternative proxy for complexity, we use the number of legal moves that can be executed in a given position. As an alternative proxy for uncertainty associated with move alternatives we use the standard deviation of the evaluation of the six best continuation moves suggested by the chess engine in a given position.} The results are also robust to the estimation of an extended specification that includes controls for decision times and decision quality in previous moves to control for a potential correlation within games (see Table \ref{tab:time_spent_robust_prev_move} in the Appendix).

%(Dainis: Ich habe etwas andere measures verwendet, fuer remaining time ob es in der Zeitnote Phase vor Zug 40 ist und fuer Complexity das Restricted Engine Mass aus dem anderen Paper , dieses alternative Mass fuer Komplexitaet hatte ich nicht mehr fuer diesen Datensatz berechnet.)

\paragraph{Decision Quality.} The second set of results refers to the quality of decisions
and its association with decision speed. Table \ref{tab:performance} presents the results from multivariate regressions
with move quality, as proxied by a binary indicator of whether the best move is played, as dependent
variable. All specifications include controls for (interacted) player-game fixed effects, current
positions, and fixed effects for the number of previously played moves.
Column (1) presents the empirical results for the association between decision quality and speed. The findings
show that faster decisions (when less time is spent on the decision) are associated with
greater decision quality (a higher propensity of making the best possible move).
The results in Columns (2)-(4) refer to the relation between decision quality and the
determinants of decision speed in isolation. For each of the determinants, the sign of the respective
coefficient corresponds to a negative association between the implied time for a decision and decision quality,
consistent with the results of Table \ref{tab:time_spent}.

\begin{table}[!t]
	\begin{center}
		\caption{Determinants of Decision Quality \label{tab:performance}}
		\footnotesize
		{
\def\sym#1{\ifmmode^{#1}\else\(^{#1}\)\fi}
\begin{tabularx}{\textwidth}{l@{\extracolsep{\fill}}cccccc}
\toprule
 &\multicolumn{6}{c}{Dependent Variable:}\\
                    &\multicolumn{6}{c}{Best Move (Dummy)}                                                                                              \\\cmidrule(lr){2-7}
                    &\multicolumn{1}{c}{(1)}         &\multicolumn{1}{c}{(2)}         &\multicolumn{1}{c}{(3)}         &\multicolumn{1}{c}{(4)}         &\multicolumn{1}{c}{(5)}         &\multicolumn{1}{c}{(6)}         \\
\midrule
\emph{Decision Time}&                     &                     &                     &                     &                     &                     \\
Time spent on move (min.)&     -0.0227\sym{***}&                     &                     &                     &     -0.0164\sym{***}&     -0.0189\sym{***}\\
                    &    (0.0006)         &                     &                     &                     &    (0.0005)         &    (0.0010)         \\
\emph{Time Budget}  &                     &                     &                     &                     &                     &                     \\
Remaining time (min.)&                     &     -0.0019\sym{***}&                     &                     &      0.0006\sym{***}&      0.0007\sym{***}\\
                    &                     &    (0.0002)         &                     &                     &    (0.0002)         &    (0.0002)         \\
\emph{Complexity}   &                     &                     &                     &                     &                     &                     \\
N Mega-nodes computed&                     &                     &     -0.0088\sym{***}&                     &     -0.0056\sym{***}&     -0.0050\sym{***}\\
                    &                     &                     &    (0.0005)         &                     &    (0.0004)         &    (0.0004)         \\
\emph{Evaluation Gap}&                     &                     &                     &                     &                     &                     \\
Distance second best move (log)&                     &                     &                     &      0.2132\sym{***}&      0.1931\sym{***}&      0.1839\sym{***}\\
                    &                     &                     &                     &    (0.0036)         &    (0.0034)         &    (0.0035)         \\
\emph{Interactions} &                     &                     &                     &                     &                     &                     \\
Time spent &                     &                     &                     &                     &                     &     -0.0002\sym{**} \\
\hspace{0.3cm}  $\times$ N Mega-nodes computed                  &                     &                     &                     &                     &                     &    (0.0001)         \\
Time spent &                     &                     &                     &                     &                     &      0.0134\sym{***}\\
\hspace{0.3cm}  $\times$ Dist. second best move                  &                     &                     &                     &                     &                     &    (0.0020)         \\
\midrule
Player-Game Fixed Effects&         Yes         &         Yes         &         Yes         &         Yes         &         Yes         &         Yes         \\
Number Move Fixed Effects&         Yes         &         Yes         &         Yes         &         Yes         &         Yes         &         Yes         \\
Control Evaluation Position&         Yes         &         Yes         &         Yes         &         Yes         &         Yes         &         Yes         \\
Move Observations   &       80359         &       80359         &       80359         &       80359         &       80359         &       80359         \\
Game Observations   &        1592         &        1592         &        1592         &        1592         &        1592         &        1592         \\
\bottomrule
\end{tabularx}
}

		\parbox{\textwidth}{
			\footnotesize \emph{Note: }{OLS estimates. The variable \emph{Distance second best move} is computed as $ln(d+1)$ where $d$ is the absolute difference between the evaluation of the best and the second best move in terms of pawn units as given by the chess engine. The evaluation of the current position is controlled for using two dummy variables indicating whether the current position is evaluated as better ($>0.5$ pawn units) or worse ($<-0.5$ pawn units) for the player to move.
				Standard errors are clustered on the game level. $^{*}$: $p<0.1$, $^{**}$: $p<0.05$, $^{***}$: $p<0.01$. }}
	\end{center}
\end{table}

Column (5) presents the results for a specification with decision time and the different proxies. The results
reveal a more subtle picture. In this specification, the negative association between decision time and decision quality
is confirmed. In addition, the results reveal independent effects of the time budget, complexity and the evaluation gap
on decision quality. In particular, once conditioning on decision time, the sign of the effect of remaining
time budget turns positive and significant. This is consistent with the hypothesis that, conditional on the decision time,
a larger time budget is associated with higher decision quality. This suggests an independent effect
of lower cogitation costs, potentially due to less emotional stress or cognitive load.
This effect is unlikely to be due to physiological factors related to, e.g., better concentration
when fatigue is lower, because such factors are captured by the control for the number of previously played moves.
Conditional on decision time, greater complexity is associated with a lower quality of the decision on average.
Likewise, greater uncertainty (as proxied by a smaller evaluation distance) is associated with a lower
decision quality. The distinct influence of response time, remaining time, and difficulty (in terms of
complexity and the evaluation gap) provide support for concerns about difficulty as a potential confounder
in the relation between response times and performance \citep[see ][]{krajbich/etal:2015,spiliopoulos2018a}.

\begin{figure}[!t]
	\begin{center}
		\begin{subfigure}{0.45\textwidth}
			\includegraphics[width=1\textwidth]{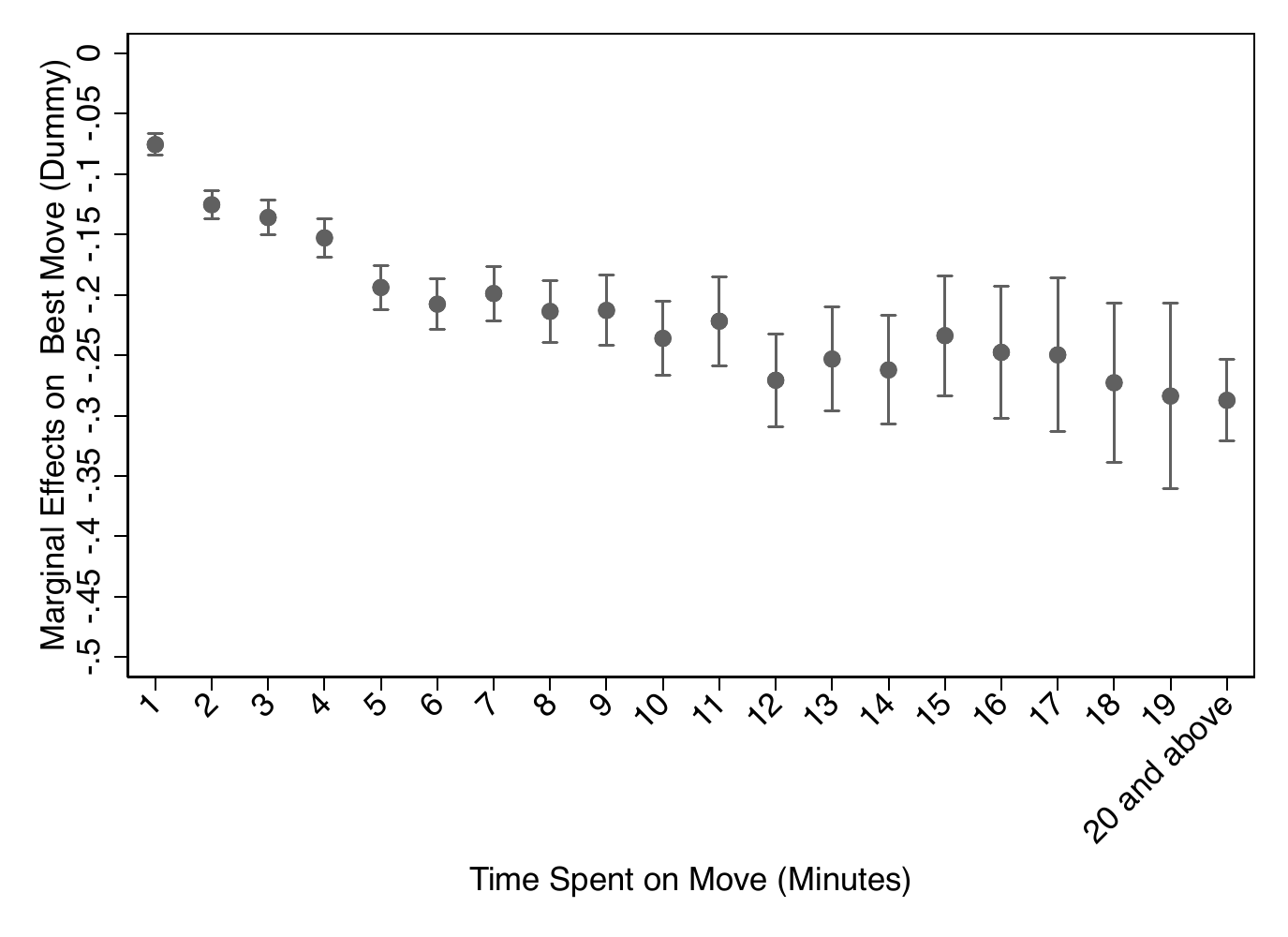}
			\caption{Decision Speed: Time Spent on Move}
		\end{subfigure}
		\begin{subfigure}{0.45\textwidth}
			\includegraphics[width=1\textwidth]{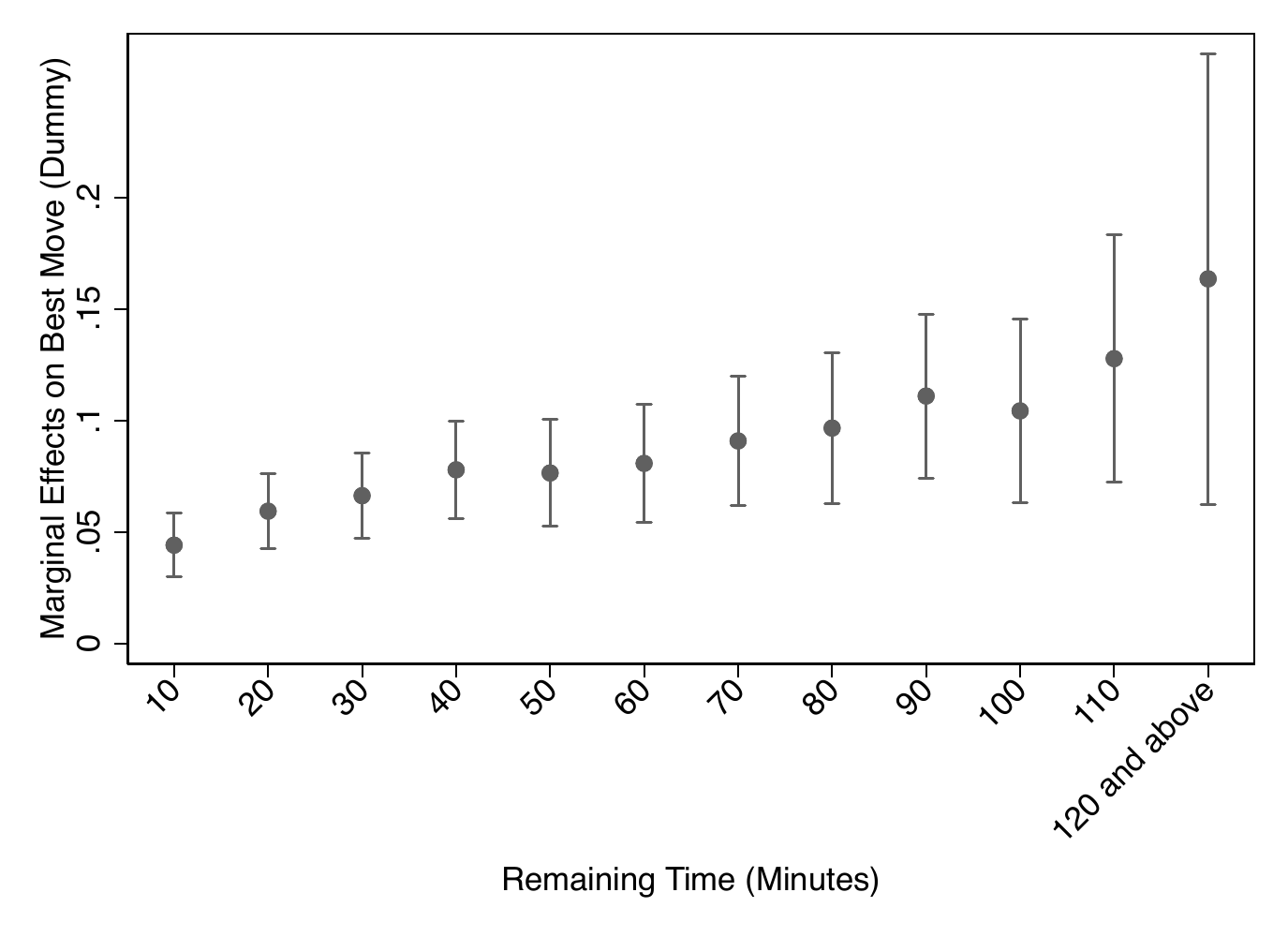}
			\caption{Time Budget: Remaining Time}
		\end{subfigure}
	\vspace{0.5cm}
	
		\begin{subfigure}{0.45\textwidth}
			\includegraphics[width=1\textwidth]{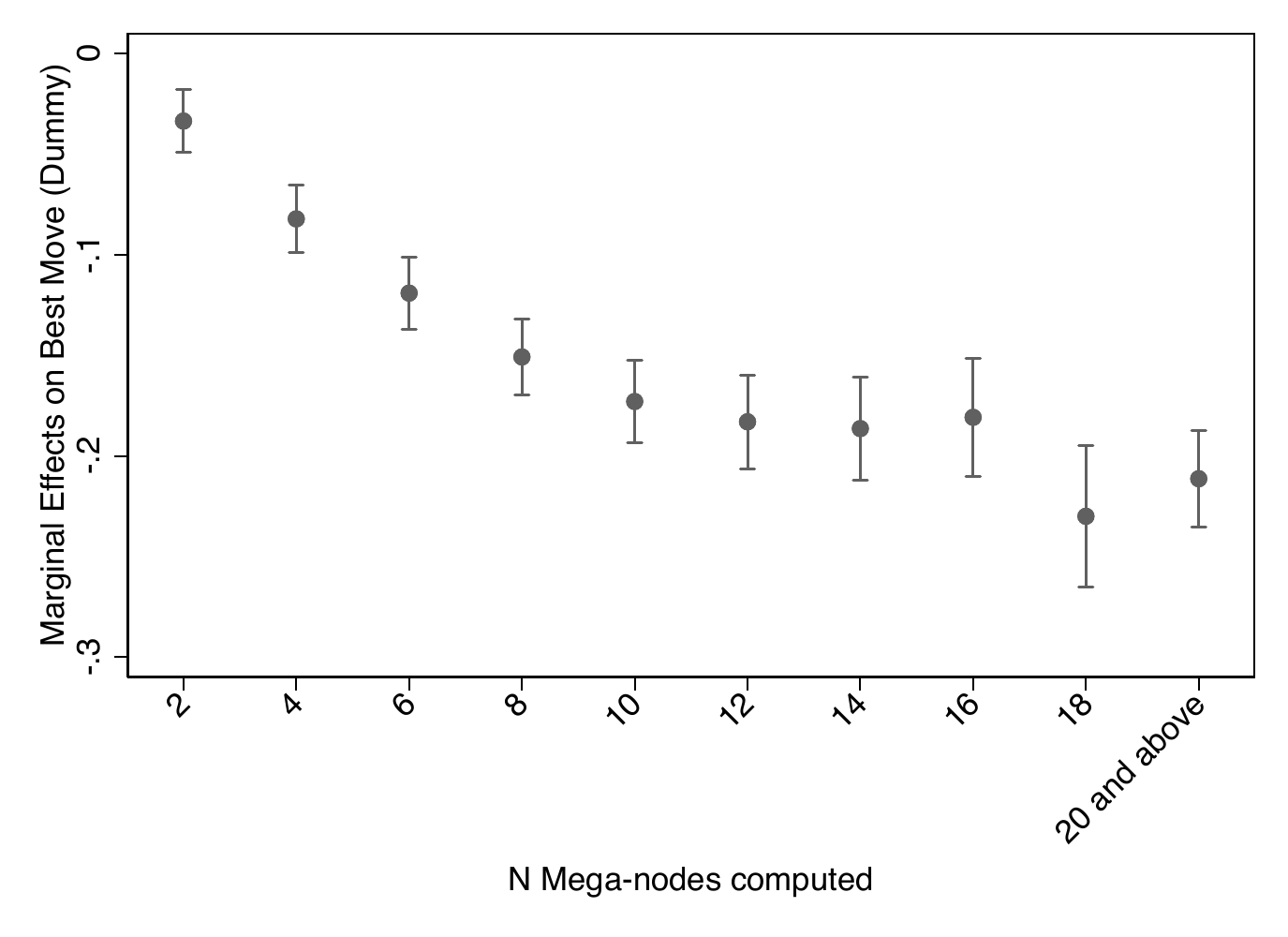}
			\caption{Complexity: Seconds to reach fixed search depth}
		\end{subfigure}
		\begin{subfigure}{0.45\textwidth}
			\includegraphics[width=1\textwidth]{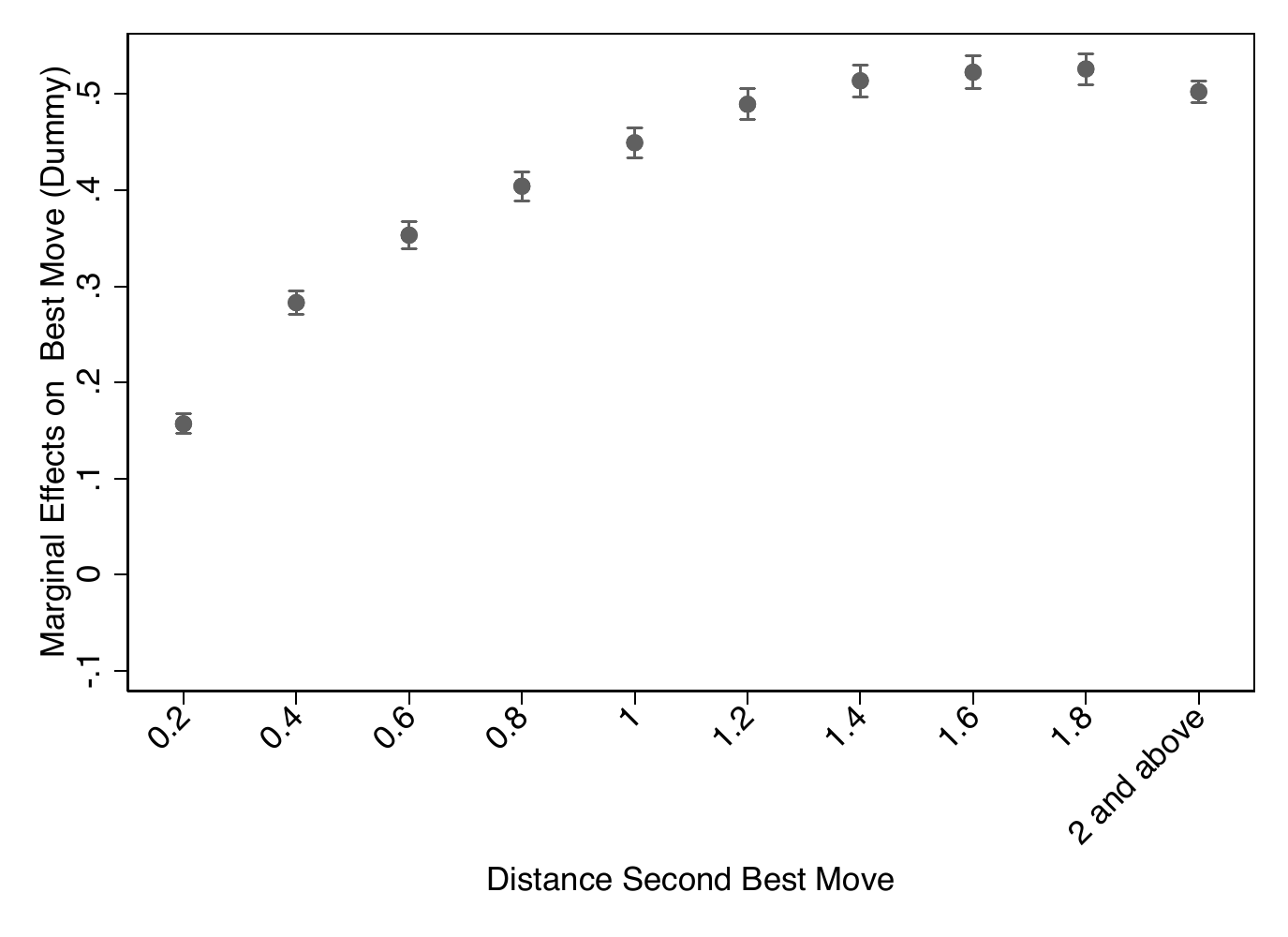}
			\caption{Evaluation Gap: Distance second best move}
		\end{subfigure}
		
		\caption{Determinants of Decision Quality - Flexible Specifications \label{fig:performance conditional}}
		\vspace{0.4cm}
		\parbox{15cm}{
			\footnotesize \emph{Note:}{ OLS results of more flexible specifications. Dependent variable is Best Move (dummy). All  graphs are based on estimates of flexible
                dummy-specifications of variables from the same regression model that includes same
                controls and fixed effects as the specifications in Table \ref{tab:performance}.
				Variables depicted on the horizontal axis are split into equally spaced intervals. Dots report point estimates and whiskers report 95\% confidence	intervals.
				Panel (a): Decision Speed measured in time spent on move in minutes; omitted reference category: time spend $<0.5$ minutes.
				Panel (b): Remaining Time measured in minutes, reference category is Remaining Time $<5$ minutes.
				Panel (c): omitted reference category: Mega-nodes computed to reach fixed search depth $<1$.
				Panel (d): Distance second best move is computed as $ln(d+1)$ where $d$ is the absolute difference between the
				evaluation of the best and the second best move in terms of pawn units as given by the chess engine;
				omitted reference category is Distance second best move $<0.1$.
		}}
	\end{center}
\end{figure}

Figure \ref{fig:performance conditional} presents corresponding non-linear relationships between decision quality and decision speed,
as well as the proxy variables for the model parameters.

Column (6) of Table \ref{tab:performance} presents results for an extended specification that includes interaction terms between time
spent and the proxies for the difficulty of a move (in terms of complexity and the uncertainty that is associated with a smaller evaluation gap).
While the main effects remain qualitatively and quantitatively
almost unchanged, the interaction terms are significant. This suggests that the effects of
complexity and uncertainty interact with decision time in determining the quality of decisions.
In particular, the negative effect of longer decision times on decision quality is amplified by greater complexity,
implying that longer cogitation does not necessarily attenuate the detrimental effect of complexity on decision quality.
Similarly, the negative effect of longer decision times is aggravated the greater the uncertainty about the best move
(in terms of a smaller distance in the evaluation of the best and second-best moves). This suggests
that longer cogitation does not help in finding the best move in complex settings or when priors about
the best decision are diffuse. In such situations, intuition and experience seem to play an important role for
the determination of the optimal decision.

The main results are robust to using alternative measures for the cost of deliberation, complexity
and the evaluation gap (see Table \ref{tab:performance_robust} in the Appendix). They are also robust to
including controls for measures of decision times and decision quality in previous moves
to account for potential correlation within games (see Table \ref{tab:performance_robust_prev_move} in the Appendix)
and when using a continuous measure for move quality along the intensive margin
of move quality in terms of pawn units (see Table \ref{tab:performance_intensive_margin}
in the Appendix).
The coefficients of the interaction terms of decision time with complexity and uncertainty
reveal opposite signs when considering decision quality along the intensive margin as dependent variable,
however (see Table \ref{tab:performance_intensive_margin} in Appendix).
%\footnote{\sout{This possibly reflects a mechanical effect,
%since if the distance to the second best move is larger, errors have a larger magnitude by construction,
%which implies a mechanically more negative coefficient.}}

%==============================================================================%
\section{Discussion}\label{sec: conclusion}
%==============================================================================%

The evidence presented in the previous section presents two sets of results on the
determinants of decision speed and the relation between decision speed and decision quality
in a real-world setting of complex and cognitively demanding decisions, using
high-frequency data from professional chess.
The first set of results documents that greater time pressure leads to faster decisions, while more difficult
decisions and decisions that are associated with greater a priori uncertainty about the best alternative
lead to slower decisions. The second set of results documents a positive relationship between decision speed and decision quality, using
an empirical strategy for decision quality that is based on a benchmark that utilizes the artificial intelligence
embodied in chess engines.

These findings are consistent with the predictions of drift-diffusion models with uncertain evaluations of decision alternatives.
While these models have been considered in the context of simple choices (like choosing between two different candy bars), our results
document that the predictions extend to complex strategic decision contexts and provide evidence on the underlying mechanisms. In drift-diffusion models, a sub-class of bounded accumulation models, noisy information is accumulated until a response criterion (a ``boundary'') is reached. Once this criterion, which reflects the amount of evidence needed to make a decision, is reached, the decision is made accordingly \citep[see, e.g.,][]{Bogacz/etal:2006,Ratcliff/etal:2015}.
Recent work has shown that bounded accumulation models have an analogue representation as a random utility model of discrete choice \citep{Webb:2019}. An important implication of this equivalence is that the observed distribution of decisions (or preferences) crucially depends on decision times.

In the present context, the decision regarding the optimal continuation move for a given configuration on the board corresponds to
the situation considered in drift-diffusion models with uncertain evaluations. Information acquisition occurs
in the form of cogitation, which is modelled as a diffusion process with drift
and random stimuli, and a decision about a move is made when the cogitation process reaches decision a boundary.\footnote{Most
existing treatments of the drift-diffusion model consider choices between
two alternatives. The characterization below extends to all choice pairs in the set of feasible choices
\citep[see][for details]{Fudenberg/etal:2020}. Recent work has generalized the model to multiple alternatives
\citep[see, e.g.,][]{Roxin:2019} without affecting the main implications for the purposes of this study.
Hence, for simplicity and concreteness, let the decision be represented either
by the best possible move, or by any other move, i.e., the decision is modeled as binary choice $i\in X$
with $X=\left\{x,y\right\}$. The diffusion process is modeled as $Z_t=\delta t + \alpha B_t$  in continuous
time $t$ with drift $\delta$ and a
stimulus process $B_t$ that is a Brownian motion with a parameter $\alpha$.
With binary choices, there are two boundaries, $b^x$ or $b^y$. With symmetric boundaries, $x$ is chosen
if $b\leq Z_t$ and $y$ if $-b\geq Z_t$.}
As consequence, the observation that a move is made is associated with a realization of decision time and of decision quality,
which is measured as the probability of making the best possible move in a given configuration conditional
on making the decision at a certain time.
%\footnote{With $p^i(t)$ denoting the probability of making the optimal
%move in a configuration $i$ at time $t$, the cdf of the marginal distribution of decision times (intuitively,
%the probability that a decision has been made by $t$) can be characterized as $F(t)$.}
Uncertainty enters in terms of the evaluations of the choice alternatives. The
decision maker has a prior belief about the evaluations, but is uncertain about the best move,
either due to the analytical complexity of the problem or due to the strategic uncertainty.
Hence, the decision maker incurs costs
over time while observing (independent) signals about the true evaluation (which also have the form of drift
plus random stimuli that follow a Brownian motion). Since the decision depends on the relative evaluation
of the choice alternatives, the evaluation difference is a sufficient statistic for the decision.
%, $u(x)$ and $u(y)$,  such that the drift process corresponds to $\delta=u(x)-u(y)$.
Hence, the larger the perceived difference in evaluations the more informative is the signal (the larger
is the drift towards the boundaries). It has been shown that the decision boundaries
vary with decision time in potentially non-monotone ways, but the decision quality decreases with decision
time when aggregating many decisions of a decision maker who on average has correct priors about the difference in the
evaluation of the respective choice alternatives \citep[see][for details]{Fudenberg/etal:2018,Fudenberg/etal:2020}.
In this setting, the optimal choice can be characterized as the maximization of the posterior belief about
the best possible move, subject to the cost associated with cogitation per instant of time. Equivalently, the
decision problem can be characterized as the minimization of the regret of a wrong choice and of the sampling
cost.\footnote{See Proposition 2 of \citet{Fudenberg/etal:2018}.}

The probability of making the best possible move and the associated decision time are then affected by three mechanisms.
The first is related to cogitation and information acquisition and implies that stopping later to make a move
is associated with more information, which tends to increase the probability of choosing the optimal move.
At the same time, two selection effects work in the opposite direction. First, optimal decisions are associated
with earlier endogenous stopping when signals are strong and thus more informative (i.e., they exhibit a larger drift
and/or a less noisy stimulus process). Second, endogenous stopping occurs earlier in trials where subjects
experienced a faster understanding of the optimal behavior (which is equivalent to a more discriminatory prior
and thus more informative signals). This selection effect only emerges in drift-diffusion models
in which the evaluations associated with choices (i.e., their quality) are uncertain.
%Theorem 1 of \citet{Fudenberg/etal:2018} shows that the relation between speed and accuracy is proportional to the slope of the boundary, whose sign is ambigous, since (i) stopping later is associated with more information and thus more likely an optimal move, but (ii) a selection due to endogenous stopping when the signal hits the boundary, which happens more frequently when signals were strong and thus informative.

The characterization of the optimal stopping time in such a setting implies that with a higher cogitation cost, decisions are made
earlier, and that the probability of making the best move is non-increasing in cogitation cost conditional on decision
time. % \citep[Theorem 4 and Corollary 1 of][]{Fudenberg/etal:2018}.
This prediction follows from a slow-down of updating as the value of cogitation declines with the time spent on cogitating over a decision.
It is consistent with the finding of greater remaining time (i.e., lower time pressure) being associated with slower decisions
(see Table \ref{tab:time_spent} and Figure \ref{fig:speed conditional}(a)) and, conditional on decision time,
with higher decision quality (see Table \ref{tab:performance} (column 5) and Figure \ref{fig:performance conditional}(b)).
The decision time also depends on the relative strength of the drift process and the noisiness of the signal, which are
proxied by complexity and the evaluation gap between move alternatives, respectively. More informative signals
are predicted to imply faster drift and/or less noise, and thus, faster and better decisions.
Again, this is in line with the findings for complexity and the evaluation gap for decision speed
(see Table \ref{tab:time_spent} and  Figures \ref{fig:speed conditional}(b) and (c), respectively) and
decision quality (see Table \ref{tab:performance} and Figure \ref{fig:performance conditional}(c) and (d), respectively).
Moreover, in this setting, the two selection effects dominate, implying that decision
quality decreases with decision time. %\citep[see Theorem 2 in][]{Fudenberg/etal:2018}.
This prediction is consistent with the negative association between time spent on a move and the quality of the move
(see Table \ref{tab:performance} and Figure \ref{fig:performance conditional}(a)).
These theoretical predictions are robust to asymmetric priors, endogenously divided attention on the two decision alternatives,
or can be extended to non-linear cogitation (in particular increasing) cost \citep[see][for details]{Fudenberg/etal:2018}.
In sum, the empirical results regarding decision time and decision quality are consistent with the predictions of
uncertain-difference drift diffusion models.

Additional findings suggest that spending more time on a decision aggravates the negative effect of greater complexity,
while spending more time on a decision when the uncertainty about the best decision is lower improves decision quality.
These findings might be an indication of additional behavioral aspects that determine
decision making as reflected in more elaborate models, such as dual-process diffusion models that
account for alignment or conflict in the inference about the optimal decisions from a heuristic
model of decision making and a maximizing model of decision making \citep[see, e.g.,][]{alosferrer2018}.
The analysis here focused on the overall patterns of the relation between speed and decision quality in the context of
endogenous timing of decisions and abstracted from analyzing systematic heterogeneity across individuals or
the implications of strategic interactions.
For instance, to the extent that cognitive processes involved with decision
making are affected by physiological factors, age might play a crucial role for
cognitive performance \citep[see, e.g.,][]{Strittmatter/Sunde/Zegners:2020a}. Similarly,
experience might interfere systematically with the way complexity or uncertainty shape
decisions. A natural next step in the research agenda is to apply the
methodology developed here to investigate the influence of individual-specific factors for the behavioral
mechanisms in more detail. Disentangling the role of strategic interactions in shaping behavior constitutes another
interesting direction for future research.

%abstracted from strategic considerations. A natural next step in the research agenda is to apply the
%methodology developed here to investigate the respective behavioral mechanisms in more detail and
%to explicitly consider strategic interactions.

%==============================================================================%
%   B I B L I O G R A P H Y
%==============================================================================%
\newpage
\normalsize \renewcommand{\baselinestretch}{1}\normalsize

\bibliographystyle{ecta}
\bibliography{references}
%==============================================================================%

%==============================================================================%
%    A P P E N D I X
%==============================================================================%
\clearpage
\appendix % \begin{appendix} \end{appendix}

\renewcommand{\thetable}{A\arabic{table}}
\setcounter{table}{0}

\renewcommand{\thefigure}{A\arabic{figure}}
\setcounter{figure}{0}

%==============================================================================%
\section*{Appendix with Supplementary Material \newline For Online Publication}
%==============================================================================%

%%==============================================================================%
%\subsection*{Additional Figures}
%%==============================================================================%
%
%
%
%
%\clearpage
%\newpage
%

%==============================================================================%
\subsection*{Additional Tables}
%==============================================================================%

\begin{table}[h!]
	\begin{center}	
	\caption{Tournaments included in Dataset\label{tab: tournaments}}
	\footnotesize
		\begin{tabular}{lr}  	\hline \hline Tournament & Number Moves \\  
 \hline	\multicolumn{2}{c}{National Championships}  \\
  \hline Armenian Championship (2016 - 2017) & 3994 \\  
Belarusian Championship (2017) & 413  \\   
Dutch Championship (2015) & 1239  \\   
French Championship (2016) & 503  \\  
Icelandic Chess Championship (2016) & 102  \\   
Kazakhstan Championship (2016) & 103  \\   
Paraguay Championship (2016) &  44  \\   
Polish Championship (2016) & 2273 \\   
Russian Championship (2014, 2016) & 5394 \\  
Russian Women's Championship (2016) & 22 \\  
Swedish Championship (2016) & 256  \\  
Swiss Championship (2016) &  60  \\   
Ukrainian Championship (2016) & 1401  \\  
	\hline \multicolumn{2}{c}{World Chess Federation Tournaments}  \\
 \hline FIDE Grand Prix - Baku (2014) & 2334  \\   
FIDE Grand Prix - Khanty-Mansiysk (2015) & 1986 \\  
FIDE Grand Prix - Tashkent (2014) & 2610  \\  
FIDE Grand Prix - Tbilisi (2015) & 2295  \\   
FIDE Women Grand Prix - Monte Carlo (2015) & 1454  \\   
FIDE Women Grand Prix - Theran (2016) & 1019  \\ 
FIDE Women Grand Prix - Batumi (2016) & 447 \\  
	\hline \multicolumn{2}{c}{Invitational Tournaments} \\ 
 \hline Barcelona Masters (2016) & 492  \\ 
Batavia Chess Tournament (2016) &  42   \\ 
Dortmund Sparkassen Chess Meeting (2015 - 2016) & 3028   \\  
GRENKE Chess Classic (2014) & 552   \\   
Karpov Poikovsky Tournament (2016) & 1749  \\  
Khanty-Mansiysk Women's Grand Prix (2016) &  84   \\   
Lake Sevan (2016) & 2409   \\  
 Lev Gutman Birthday Jubilee Tournament (2015) &  78  \\   
London Chess Classic (2014 and 2016) & 1971   \\   
Margaryan Memorial (2017) & 165   \\  
Norway Chess (2014 - 2016) & 4536  \\
Poikovsky (2015) & 1964   \\  
Shamkir Chess (2014 - 2015) & 3715  \\   
Sigeman \& Co (2014) & 106   \\  
Sinquefield Cup (2016) & 1866   \\  
Suleymanpasa GM Tournament (2016) & 216   \\   
Tal Memorial (2016) & 2295   \\   
Tata Steel Challengers (2016 - 2017) & 5595   \\   
Tata Steel Masters (2014, 2016 - 2017) & 9052  \\  
U.S. Junior Championship (2016) & 112   \\  
UKA German Masters (2015) &  98  \\   
US.Championships (2015 - 2016) & 5925   \\   
Vidmar Memorial (2016) & 2568   \\    \hline
Sum & 80359   \\   
\end{tabular}

	\end{center}
\footnotesize{\emph{Note: } This table shows an overview of tournaments that are included in our dataset. The tournaments include all elite single round-robin tournaments played in 2014 - 2017. Games with likely errounous decision times such as negative decision times or decision times of several hours for a single move were excluded. Further, the data is restriced to games were both players have at least an ELO number of 2500. The data are taken from \url{chess24.com}.}

\end{table}

\begin{table}[h!]
	\caption{Descriptive Statistics --  Move Level\label{tab: descriptive statistics}}
	\footnotesize
	%\tiny
	\begin{center}	
{
\def\sym#1{\ifmmode^{#1}\else\(^{#1}\)\fi}
\begin{tabular}{l*{1}{ccccc}}
\toprule
                    &       Moves&        Mean&   Std. Dev.&         Min&         Max\\
\midrule
\emph{Decision Quality}&            &            &            &            &            \\
Best move (dummy)   &      80,359&       0.565&       0.496&           0&           1\\
Move quality (log-modulus)&      80,359&      -0.207&       0.406&      -5.784&           0\\
\emph{Decision Time}&            &            &            &            &            \\
Time spent on move (min.)&      80,359&       2.712&       4.241&           0&       63.67\\
\emph{Time Budget}  &            &            &            &            &            \\
Remaining time (min.)&      80,359&       32.59&       26.49&      0.0167&       140.8\\
Time pressure phase before move 40&      80,359&       0.240&       0.427&           0&           1\\
\emph{Complexity}   &            &            &            &            &            \\
N Mega-nodes computed&      80,359&       6.517&       7.610&    0.000873&       241.8\\
N moves possible    &      80,359&       29.95&       11.78&           2&          67\\
\emph{Evaluation Gap}&            &            &            &            &            \\
Distance second best move (log)&      80,359&       0.501&       0.786&           0&       5.957\\
Standard deviation evaluation 6 best moves&      80,359&       5.235&       23.40&           0&       267.0\\
\bottomrule
\end{tabular}
}

\end{center}
\vspace{0.2cm}
\footnotesize{\emph{Note: } Descriptive statistics for the baseline sample.
	The variable $Move~Quality$ is calculated as $sign(d)\cdot \ln(|d| + 1)$, where $d$ measures the difference in the evaluation of the move played relative to the best possible move in pawn units.  The variable \emph{N moves possible} represents the number of possible legal moves in a given configuration. The variable \emph{Distance second best move} is computed as $ln(d+1)$ where $d$ is the absolute difference between the evaluation of the best and the second best move in terms of pawn units as given by the chess engine. The variable \emph{Standard deviation evaluation 6 best moves} is calculated as the standard deviation of the evaluation of the chess positions according to the chess engine after each of the six best moves in a chess position would have been played in a given position.}
\end{table}

\begin{table}[!t]
\begin{center}
\caption{Determinants of Decision Time - Robustness - Alternative Measures\label{tab:time_spent_robust}}
\footnotesize
{
\def\sym#1{\ifmmode^{#1}\else\(^{#1}\)\fi}
\begin{tabularx}{\textwidth}{l@{\extracolsep{\fill}}ccccc}
\toprule
 &\multicolumn{4}{c}{Dependent Variable:}\\
                    &\multicolumn{4}{c}{Time spent on move (minutes)}                                       \\\cmidrule(lr){2-5}
                    &\multicolumn{1}{c}{(1)}         &\multicolumn{1}{c}{(2)}         &\multicolumn{1}{c}{(3)}         &\multicolumn{1}{c}{(4)}         \\
\midrule
\emph{Time Budget}  &                     &                     &                     &                     \\
Time pressure phase before move 40&     -1.3256\sym{***}&                     &                     &     -1.2581\sym{***}\\
                    &    (0.0411)         &                     &                     &    (0.0411)         \\
\emph{Complexity}   &                     &                     &                     &                     \\
N moves possible    &                     &      0.0541\sym{***}&                     &      0.0443\sym{***}\\
                    &                     &    (0.0021)         &                     &    (0.0021)         \\
\emph{Evaluation Gap}&                     &                     &                     &                     \\
Standard deviation evaluation 6 best moves&                     &                     &     -0.0122\sym{***}&     -0.0080\sym{***}\\
                    &                     &                     &    (0.0006)         &    (0.0005)         \\
\midrule
Player-Game Fixed Effects&         Yes         &         Yes         &         Yes         &         Yes         \\
Control Move Number &         Yes         &         Yes         &         Yes         &         Yes         \\
Control Evaluation Position&         Yes         &         Yes         &         Yes         &         Yes         \\
Move Observations   &       80359         &       80359         &       80359         &       80359         \\
Game Observations   &        1592         &        1592         &        1592         &        1592         \\
\bottomrule
\end{tabularx}
}

\parbox{\textwidth}{\footnotesize \emph{Note: }{OLS estimates. The variable \emph{Time pressure phase before move 40} is an alternative proxy for the time cost of cogitation and represents a dummy variable indicating whether the move number is in the range between move 30 and move 39. At move 40, each player receives additional time onto his time budget, therefore the moves before 40 are typically characterized as the time-pressure phase in chess. Since this does not allow to include fixed effects for the move number at the same time, all specification include a linear control variable for the move number. The variable \emph{N moves possible} is an alternative proxy for complexity and represents the number of possible legal moves in a given configuration. A higher number of possible moves is typically associated with higher complexity. The variable \emph{Standard deviation evaluation 6 best moves} is an alternative proxy for uncertainty of the prior and represents a measure of the standard deviation of the evaluation of the chess positions according to the chess engine after each of the six best moves in a chess position would have been played in a given position. The evaluation of the current position is controlled for using two dummy variables indicating whether the current position is evaluated as better ($>0.5$ pawn units) or worse ($<-0.5$ pawn units) for the player to move. Standard errors are clustered on the game level. $^{*}$: $p<0.1$, $^{**}$: $p<0.05$, $^{***}$: $p<0.01$.}}
\end{center}
\end{table}

\begin{landscape}
\begin{table}[!t]
\begin{center}
\caption{Determinants of Decision Time - Robustnesss - Previous Own Moves and Opponent
	\label{tab:time_spent_robust_prev_move}
}
\footnotesize
{
\def\sym#1{\ifmmode^{#1}\else\(^{#1}\)\fi}
\begin{tabularx}{1.25\textwidth}{l@{\extracolsep{\fill}}cccccc}
\toprule
 &\multicolumn{5}{c}{Dependent Variable:}\\
                    &\multicolumn{5}{c}{Time Spent on Move (minutes)}                                                             \\\cmidrule(lr){2-6}
                    &\multicolumn{1}{c}{(1)}         &\multicolumn{1}{c}{(2)}         &\multicolumn{1}{c}{(3)}         &\multicolumn{1}{c}{(4)}         &\multicolumn{1}{c}{(5)}         \\
\midrule
\emph{Time Budget}  &                     &                     &                     &                     &                     \\
Remaining time (min.)&      0.0947\sym{***}&      0.0951\sym{***}&      0.0929\sym{***}&      0.0929\sym{***}&      0.0926\sym{***}\\
                    &    (0.0032)         &    (0.0032)         &    (0.0031)         &    (0.0031)         &    (0.0031)         \\
\emph{Complexity}   &                     &                     &                     &                     &                     \\
N Mega-nodes computed&      0.0193\sym{***}&      0.0193\sym{***}&      0.0197\sym{***}&      0.0199\sym{***}&      0.0199\sym{***}\\
                    &    (0.0028)         &    (0.0028)         &    (0.0026)         &    (0.0026)         &    (0.0026)         \\
\emph{Evaluation Gap}&                     &                     &                     &                     &                     \\
Distance second best move (log)&     -0.8294\sym{***}&     -0.8291\sym{***}&     -0.8286\sym{***}&     -0.8332\sym{***}&     -0.8296\sym{***}\\
                    &    (0.0233)         &    (0.0232)         &    (0.0228)         &    (0.0228)         &    (0.0227)         \\
\emph{Decision Time Previous Moves}&                     &                     &                     &                     &                     \\
Time spent on move (previous own move)&                     &      0.0053         &     -0.0125\sym{**} &     -0.0098\sym{**} &     -0.0087\sym{*}  \\
                    &                     &    (0.0049)         &    (0.0049)         &    (0.0050)         &    (0.0050)         \\
Time spent on move (previous opponent's move)&                     &                     &      0.1313\sym{***}&      0.1321\sym{***}&      0.1264\sym{***}\\
                    &                     &                     &    (0.0074)         &    (0.0074)         &    (0.0075)         \\
\emph{Decision Quality Previous Moves}&                     &                     &                     &                     &                     \\
Best move (previous own move)&                     &                     &                     &      0.1217\sym{***}&      0.1289\sym{***}\\
                    &                     &                     &                     &    (0.0292)         &    (0.0291)         \\
Best move (previous opponent's move)&                     &                     &                     &                     &     -0.2496\sym{***}\\
                    &                     &                     &                     &                     &    (0.0309)         \\
\midrule
Player-Game Fixed Effects&         Yes         &         Yes         &         Yes         &         Yes         &         Yes         \\
Control Move Number &         Yes         &         Yes         &         Yes         &         Yes         &         Yes         \\
Control Evaluation Position&         Yes         &         Yes         &         Yes         &         Yes         &         Yes         \\
Move Observations   &       80359         &       80359         &       80359         &       80359         &       80359         \\
Game Observations   &        1592         &        1592         &        1592         &        1592         &        1592         \\
\bottomrule
\end{tabularx}
}

\vspace{0.5cm}
\parbox{1.26\textwidth}{
	\footnotesize \emph{Note: }{ OLS estimates. Column (1) shows for comparision the main specification from table  \ref{tab:time_spent}. The evaluation of the current position is controlled for using two dummy variables indicating whether the current position is evaluated as better ($>0.5$ pawn units) or worse ($<-0.5$ pawn units) for the player to move. Standard errors are clustered on the game level. $^{*}$: $p<0.1$, $^{**}$: $p<0.05$, $^{***}$: $p<0.01$. }}
\end{center}
\end{table}
\end{landscape}

\begin{landscape}
\begin{table}[!t]
\begin{center}
\caption{Determinants of Decision Quality - Robustness - Alternative Measures\label{tab:performance_robust}}
\footnotesize
{
\def\sym#1{\ifmmode^{#1}\else\(^{#1}\)\fi}
\begin{tabularx}{1.25\textwidth}{l@{\extracolsep{\fill}}cccccc}
\toprule
 &\multicolumn{6}{c}{Dependent Variable:}\\
                    &\multicolumn{6}{c}{Best Move (Dummy)}                                                                                              \\\cmidrule(lr){2-7}
                    &\multicolumn{1}{c}{(1)}         &\multicolumn{1}{c}{(2)}         &\multicolumn{1}{c}{(3)}         &\multicolumn{1}{c}{(4)}         &\multicolumn{1}{c}{(5)}         &\multicolumn{1}{c}{(6)}         \\
\midrule
Time spent on move (min.)&     -0.0215\sym{***}&                     &                     &                     &     -0.0207\sym{***}&     -0.0222\sym{***}\\
                    &    (0.0006)         &                     &                     &                     &    (0.0006)         &    (0.0024)         \\
\emph{Time Budget}  &                     &                     &                     &                     &                     &                     \\
Time pressure phase before move 40&                     &     -0.0096\sym{*}  &                     &                     &     -0.0432\sym{***}&     -0.0430\sym{***}\\
                    &                     &    (0.0051)         &                     &                     &    (0.0051)         &    (0.0050)         \\
\emph{Complexity}   &                     &                     &                     &                     &                     &                     \\
N moves possible    &                     &                     &     -0.0057\sym{***}&                     &     -0.0036\sym{***}&     -0.0037\sym{***}\\
                    &                     &                     &    (0.0002)         &                     &    (0.0002)         &    (0.0003)         \\
\emph{Evaluation Gap}&                     &                     &                     &                     &                     &                     \\
Standard deviation evaluation 6 best moves&                     &                     &                     &      0.0023\sym{***}&      0.0017\sym{***}&      0.0017\sym{***}\\
                    &                     &                     &                     &    (0.0001)         &    (0.0001)         &    (0.0001)         \\
\emph{Interactions} &                     &                     &                     &                     &                     &                     \\
N possible moves $\times$ Time spent&                     &                     &                     &                     &                     &      0.0000         \\
                    &                     &                     &                     &                     &                     &    (0.0001)         \\
Standard deviation evaluation $\times$ Time spent&                     &                     &                     &                     &                     &     -0.0000         \\
                    &                     &                     &                     &                     &                     &    (0.0000)         \\
\midrule
Player-Game Fixed Effects&         Yes         &         Yes         &         Yes         &         Yes         &         Yes         &         Yes         \\
Control Move Number &         Yes         &         Yes         &         Yes         &         Yes         &         Yes         &         Yes         \\
Control Evaluation Position&         Yes         &         Yes         &         Yes         &         Yes         &         Yes         &         Yes         \\
Move Observations   &       80359         &       80359         &       80359         &       80359         &       80359         &       80359         \\
Game Observations   &        1592         &        1592         &        1592         &        1592         &        1592         &        1592         \\
\bottomrule
\end{tabularx}
}

\vspace{0.5cm}
\parbox{1.26\textwidth}{
	\footnotesize \emph{Note: }{OLS estimates. The variable \emph{Time pressure phase before move 40} is an alternative proxy for the time cost of cogitation, $c$, and represents a dummy variable indicating whether the move number is in the range between move 30 and move 39. At move 40, each player receives additional time onto his time budget, therefore the moves before 40 are typically characterized as the time-pressure phase in chess. Since this does not allow to include fixed effects for the move number at the same time, all specification include a linear control variable for the move number. The variable \emph{N moves possible} is an alternative proxy for complexity and represents the number of possible moves in the current configuration. The variable \emph{Standard deviation evaluation 6 best moves} is an alternative proxy for uncertainty of the prior and represents a measure of the standard deviation of the evaluation of the chess positions according to the chess engine after each of the six best moves in a chess position would have been played in a given position. The evaluation of the current position is controlled for using two dummy variables indicating whether the current position is evaluated as better ($>0.5$ pawn units) or worse ($<-0.5$ pawn units) for the player to move.
Standard errors are clustered on the game level. $^{*}$: $p<0.1$, $^{**}$: $p<0.05$, $^{***}$: $p<0.01$.}}
\end{center}
\end{table}
\end{landscape}

\begin{landscape}
\begin{table}[!t]
\begin{center}
\caption{Determinants of Decision Quality - Robustnesss - Previous Own Moves and Opponent
	\label{tab:performance_robust_prev_move}
}
\footnotesize
{
\def\sym#1{\ifmmode^{#1}\else\(^{#1}\)\fi}
\begin{tabularx}{1.25\textwidth}{l@{\extracolsep{\fill}}cccccc}
\toprule
 &\multicolumn{5}{c}{Dependent Variable:}\\
                    &\multicolumn{5}{c}{Best Move (Dummy)}                                                                        \\\cmidrule(lr){2-6}
                    &\multicolumn{1}{c}{(1)}         &\multicolumn{1}{c}{(2)}         &\multicolumn{1}{c}{(3)}         &\multicolumn{1}{c}{(4)}         &\multicolumn{1}{c}{(5)}         \\
\midrule
\emph{Decision Time}&                     &                     &                     &                     &                     \\
Time spent on move (min.)&     -0.0164\sym{***}&     -0.0164\sym{***}&     -0.0169\sym{***}&     -0.0170\sym{***}&     -0.0169\sym{***}\\
                    &    (0.0005)         &    (0.0005)         &    (0.0005)         &    (0.0005)         &    (0.0005)         \\
\emph{Complexity}   &                     &                     &                     &                     &                     \\
N Mega-nodes computed&     -0.0056\sym{***}&     -0.0056\sym{***}&     -0.0056\sym{***}&     -0.0056\sym{***}&     -0.0056\sym{***}\\
                    &    (0.0004)         &    (0.0004)         &    (0.0004)         &    (0.0004)         &    (0.0004)         \\
\emph{Evaluation Gap}&                     &                     &                     &                     &                     \\
Distance second best move (log)&      0.1931\sym{***}&      0.1931\sym{***}&      0.1926\sym{***}&      0.1918\sym{***}&      0.1917\sym{***}\\
                    &    (0.0034)         &    (0.0034)         &    (0.0034)         &    (0.0034)         &    (0.0034)         \\
Remaining time (min.)&      0.0006\sym{***}&      0.0005\sym{**} &      0.0005\sym{**} &      0.0005\sym{**} &      0.0005\sym{**} \\
                    &    (0.0002)         &    (0.0002)         &    (0.0002)         &    (0.0002)         &    (0.0002)         \\
Time spent on move (previous own move)&                     &     -0.0013\sym{***}&     -0.0018\sym{***}&     -0.0013\sym{***}&     -0.0014\sym{***}\\
                    &                     &    (0.0004)         &    (0.0004)         &    (0.0004)         &    (0.0004)         \\
Time spent on move (previous opponent's move)&                     &                     &      0.0036\sym{***}&      0.0037\sym{***}&      0.0039\sym{***}\\
                    &                     &                     &    (0.0004)         &    (0.0004)         &    (0.0004)         \\
\emph{Decision Quality Previous Moves}&                     &                     &                     &                     &                     \\
Best move (previous own move)&                     &                     &                     &      0.0207\sym{***}&      0.0205\sym{***}\\
                    &                     &                     &                     &    (0.0035)         &    (0.0035)         \\
Best move (previous opponent's move)&                     &                     &                     &                     &      0.0076\sym{**} \\
                    &                     &                     &                     &                     &    (0.0037)         \\
\midrule
Player-Game Fixed Effects&         Yes         &         Yes         &         Yes         &         Yes         &         Yes         \\
Number Move Fixed Effects&         Yes         &         Yes         &         Yes         &         Yes         &         Yes         \\
Control Evaluation Position&         Yes         &         Yes         &         Yes         &         Yes         &         Yes         \\
Move Observations   &       80359         &       80359         &       80359         &       80359         &       80359         \\
Game Observations   &        1592         &        1592         &        1592         &        1592         &        1592         \\
\bottomrule
\end{tabularx}
}

\vspace{0.5cm}
\parbox{1.26\textwidth}{
	\footnotesize \emph{Note: }{ OLS estimates. Column (1) shows for comparison the main specification without interactions from table  \ref{tab:performance}. The evaluation of the current position is controlled for using two dummy variables indicating whether the current position is evaluated as better ($>0.5$ pawn units) or worse ($<-0.5$ pawn units) for the player to move. Standard errors are clustered on the game level. $^{*}$: $p<0.1$, $^{**}$: $p<0.05$, $^{***}$: $p<0.01$. }}
\end{center}
\end{table}
\end{landscape}

\begin{landscape}
\begin{table}[!t]
\begin{center}
\caption{Determinants of Decision Quality - Robustness - Intensive Margin \label{tab:performance_intensive_margin}}
\footnotesize
{
\def\sym#1{\ifmmode^{#1}\else\(^{#1}\)\fi}
\begin{tabularx}{1.25\textwidth}{l@{\extracolsep{\fill}}cccccc}
\toprule
 &\multicolumn{6}{c}{Dependent Variable:}\\
                    &\multicolumn{6}{c}{Move Quality (Log-Modulus)}                                                                                     \\\cmidrule(lr){2-7}
                    &\multicolumn{1}{c}{(1)}         &\multicolumn{1}{c}{(2)}         &\multicolumn{1}{c}{(3)}         &\multicolumn{1}{c}{(4)}         &\multicolumn{1}{c}{(5)}         &\multicolumn{1}{c}{(6)}         \\
\midrule
\emph{Decision Time}&                     &                     &                     &                     &                     &                     \\
Time spent on move (min.)&     -0.0091\sym{***}&                     &                     &                     &     -0.0103\sym{***}&     -0.0063\sym{***}\\
                    &    (0.0004)         &                     &                     &                     &    (0.0004)         &    (0.0008)         \\
\emph{Time Budget}  &                     &                     &                     &                     &                     &                     \\
Remaining time (min.)&                     &     -0.0001         &                     &                     &      0.0009\sym{***}&      0.0008\sym{***}\\
                    &                     &    (0.0002)         &                     &                     &    (0.0002)         &    (0.0002)         \\
\emph{Complexity}   &                     &                     &                     &                     &                     &                     \\
N Mega-nodes computed&                     &                     &     -0.0082\sym{***}&                     &     -0.0084\sym{***}&     -0.0089\sym{***}\\
                    &                     &                     &    (0.0006)         &                     &    (0.0006)         &    (0.0007)         \\
\emph{Evaluation Gap}&                     &                     &                     &                     &                     &                     \\
Distance second best move (log)&                     &                     &                     &     -0.0259\sym{***}&     -0.0429\sym{***}&     -0.0307\sym{***}\\
                    &                     &                     &                     &    (0.0063)         &    (0.0065)         &    (0.0065)         \\
\emph{Interactions} &                     &                     &                     &                     &                     &                     \\
Time spent $\times$ N Mega-nodes computed&                     &                     &                     &                     &                     &      0.0002\sym{**} \\
                    &                     &                     &                     &                     &                     &    (0.0001)         \\
Time spent $\times$ Dist. second best move&                     &                     &                     &                     &                     &     -0.0174\sym{***}\\
                    &                     &                     &                     &                     &                     &    (0.0028)         \\
\midrule
Player-Game Fixed Effects&         Yes         &         Yes         &         Yes         &         Yes         &         Yes         &         Yes         \\
Number Move Fixed Effects&         Yes         &         Yes         &         Yes         &         Yes         &         Yes         &         Yes         \\
Control Evaluation Position&         Yes         &         Yes         &         Yes         &         Yes         &         Yes         &         Yes         \\
Move Observations   &       80359         &       80359         &       80359         &       80359         &       80359         &       80359         \\
Game Observations   &        1592         &        1592         &        1592         &        1592         &        1592         &        1592         \\
\bottomrule
\end{tabularx}
}

\vspace{0.5cm}
\parbox{1.26\textwidth}{
	\footnotesize \emph{Note: }{ OLS estimates. The dependent variable Move Quality (Log-Modulus) is a measure of the distance of the actual move played from the optimal move stipulated by the chess engine, in terms of the log modulus transformation of pawn units, such that $Move~Quality = sign(d)\cdot \ln(|d| + 1)$, where $d$ measures the difference in the evaluation of the move played relative to the best possible move in pawn units. The evaluation of the current position is controlled for using two dummy variables indicating whether the current position is evaluated as better ($>0.5$ pawn units) or worse ($<-0.5$ pawn units) for the player to move. Standard errors are clustered on the game level. $^{*}$: $p<0.1$, $^{**}$: $p<0.05$, $^{***}$: $p<0.01$. }}
\end{center}
\end{table}
\end{landscape}

\clearpage
\end{document}